\documentclass[a4paper,11pt]{article}
\pdfoutput=1

\usepackage{jcappub} 

\usepackage[T1]{fontenc} 

\title{\boldmath A new method to quantify the effects of baryons on the matter power spectrum}

\author[]{Aurel Schneider}
\author[]{and Romain Teyssier}
\affiliation[]{Institute for Computational Science, University of Zurich, Winterthurerstrasse 190, 8057 Zurich, Switzerland}

\emailAdd{aurel@physik.uzh.ch}
\emailAdd{teyssier@physik.uzh.ch}

\abstract{Future large-scale galaxy surveys have the potential to become leading probes for cosmology provided the influence of baryons on the total mass distribution is understood well enough. As hydrodynamical simulations strongly depend on details in the feedback implementations, no unique and robust predictions for baryonic effects currently exist. In this paper we propose a {\it baryonic correction model} that modifies the density field of dark-matter-only $N$-body simulations to mimic the effects of baryons from any underlying adopted feedback recipe. The model assumes haloes to consist of 4 components: 1- hot gas in hydrostatical equilibrium, 2- ejected gas from feedback processes, 3- central galaxy stars, and 4- adiabatically relaxed dark matter, which all modify the initial dark-matter-only density profiles. These altered profiles allow to define a displacement field for particles in $N$-body simulations and to modify the total density field accordingly.

The main advantage of the {\it baryonic correction model} is to connect the total matter density field to the observable distribution of gas and stars in haloes, making it possible to parametrise baryonic effects on the matter power spectrum. We show that the most crucial quantities are the {\it mass fraction of ejected gas} and its corresponding {\it ejection radius}. The former controls how strongly baryons suppress the power spectrum, while the latter provides a measure of the scale where baryonic effects become important. A comparison with X-ray and Sunyaev-Zel'dovich cluster observations suggests that baryons suppress wave modes above $k\sim0.5$ h/Mpc with a maximum suppression of 10-25 percent around $k\sim 2$ h/Mpc. More detailed observations of the gas in the outskirts of groups and clusters are required to decrease the large uncertainties of these numbers.}

\begin{document}
\maketitle
\flushbottom

\section{Introduction}
Cosmological structure formation is driven by the dominant yet invisible dark matter (DM) component acting as a self-gravitating collisionless fluid. The subdominant baryonic component is collisional and dissipative, what makes it considerably more complicated to model. For a long time it was considered sufficient to only predict the DM clustering while assuming galaxies to trace the centres of matter over-densities. In recent years, however, results from hydrodynamical simulations suggested that baryons affect the total matter distribution up to scales large enough to become relevant for cosmology \citep{Rudd2008,vanDaalen2011}.

Upcoming galaxy and weak-lensing surveys such as {\tt DES}\footnote{\tt www.darkenergysurvey.org}, {\tt LSST}\footnote{\tt www.lsst.org/lsst}, and {\tt Euclid}\footnote{\tt sci.esa.int/euclid} have the potential to become the next leading cosmological probes \citep{Ivezic2008,Laureijs2011,Amendola2013}. They are complementary to the highly successful cosmic microwave background (CMB) measurements \citep{Ade2014}, potentially providing cosmological information from smaller scales at much later times. The success of weak lensing surveys, however, crucially depends on our understanding of how baryons influence the matter distribution, making the study of baryons a critical task for cosmology.

Numerical simulations with full hydrodynamic and radiative treatment have become increasingly popular in the last decade and major progresses have been made \citep{Borgani2011}. The fundamental difficulty of hydrodynamical simulations, with respect to the much simpler dark-matter-only simulations, is that the large-scale gas distribution is affected by small-scale events not resolved by the simulation. Active galactic nuclei (AGN) driven by black hole accretion or single supernovae explosions provide energy feedback mechanisms altering the formation of galaxies. As they are far too small to be properly included in simulations, they are usually implemented phenomenologically as {\it sub-grid processes}. These feedback mechanisms are poorly constrained by observations and severely restrict the predicability of hydrodynamical simulations for cosmology. 

Recently, several analytical approaches have been proposed to complement results from hydrodynamical simulations and to enable a more direct understanding of how baryons affect large-scale structure formation. Examples are baryonic extensions of perturbation theory \citep{Lewandowski2015,Mohammed2014b} or modifications of the halo model \citep{Semboloni2013,Fedeli2014,Fedeli2014b,Mohammed2014,Mead2015}. The downsides of these two approaches are that perturbation theory is only accurate at large scales, while the halo model has a rather limited overall accuracy.

In this paper, we propose yet another approach to include the effects of baryons on the matter density field. The {\it baryonic correction model} consists of an analytically motivated modification of dark-matter-only simulations. It assumes every halo to be made of four different matter components -- dark matter, stars, bound gas, and ejected gas -- with specific fractions and profiles that can be constrained with observations. Together, the components sum up to a total halo profile which differs from the initial DM-only profile. This allows us to define a Lagrangian mapping and displace particles in a DM-only simulation in order to recover the modified profile for every halo.

The  {\it baryonic correction model} has the advantage of providing a simple and adaptable recipe and a direct link between observations of the gaseous and stellar profiles on the one hand and the total matter density field on the other hand. For example, fraction and radius of the ejected gas can be arbitrarily varied, illustrating how the efficiency of energy feedback affects the matter distribution. The model is not supposed to replace hydrodynamical simulations but to provide a fast and controllable tool to quantify the effects of baryons on the large-scale matter distribution.

Concerning the analysis of the matter distribution, we focus on the power spectrum, the prime statistical measure of the large-scale structure. As a requirement from upcoming weak-lensing surveys, such as {\tt Euclid}, the theoretical knowledge of the matter power spectrum has to be at percent precision at wave modes below $k\sim10$ hMpc$^{-1}$. In Ref.~\citep{Schneider2015} we showed that it is challenging to obtain this level of accuracy with DM-only simulations. As for baryons, an additional modification at the ten percent level above $k\sim0.1-1$ hMpc$^{-1}$ is expected. It is therefore of prime importance to further quantify or at least parametrise the effects of baryons on the matter power spectrum.

The paper is structured as follows. In Sec.~\ref{sec:model} we introduce the model, providing details about the different model components and how they are constrained observationally. Sec.~\ref{sec:sims} is devoted to the details of modifying $N$-body simulations and shows an example case of a single cluster. We then discuss the effects on the power spectrum and define a simple fitting function for the baryonic effects in Sec.~\ref{sec:ps}, before concluding in Sec~\ref{sec:conc}.

\section{\boldmath  Mimicking the effects of baryons}\label{sec:model}
The {\it baryonic correction model} aims to mimic the effects of gas and stars on the halo profiles of cosmological $N$-body simulations. The idea is to replace the initial dark matter NFW profile with a combination of an adiabatically relaxed dark matter profile, a stellar profile, a hot gas profile, plus a component of ejected gas, where profiles and abundances are determined by observations. We start with an overview of the model, before providing details about the shapes of the profiles and the parametrisation of the different abundance fractions.

\subsection{The model}\label{modelintro}
The outcome of $N$-body simulations can be approximatively described by haloes (identified with a halo finder) in a smooth background component. Around each halo centre the density is given by the dark-matter-only ({\it dmo}) profile
\begin{equation}\label{rhodmo}
\rho_{\rm dmo}(r)=\rho_{\rm nfw}(r)+\bar\rho_{\rm bg}\,.
\end{equation}
where $\rho_{\rm nfw}$ is a truncated NFW-profile (with finite total mass $M_{\rm tot}$) and $\rho_{\rm bg}$ is the background density of non-collapsed matter, both defined in Sec.~\ref{sec:rhoi}. 

The simplified description of Eq.~\eqref{rhodmo} can be modified to account for the effects of baryons on the total density field. A model with baryons may consist of an adiabatically relaxed DM profile ($y_{\rm rdm}$), a profile of bound gas in hydrostatic equilibrium ($y_{\rm bgas}$), expelled gas due to feedback ejection ($y_{\rm egas}$), and a stellar component from the central galaxy ($y_{\rm cgal}$). All these components can be combined to obtain the final profile of the baryonic correction model ($\rho_{\rm bcm}$), which has the form
\begin{equation}\label{rhodmb}
\rho_{\rm bcm}(r)=\,f_{\rm rdm}y_{\rm rdm}(r) +f_{\rm bgas}(M)y_{\rm bgas}(r) + f_{\rm egas}(M)y_{\rm egas}(r) + f_{\rm cgal}(M)y_{\rm cgal}(r) + \bar\rho_{\rm bg},
\end{equation}
where the normalised individual profiles $y_{\chi}$ and the corresponding abundance fractions $f_{\chi}$ are described in Sec.~\ref{sec:rhof} and \ref{sec:frac}. The mass dependences of the abundance fractions comes from the fact that heating and ejection of gas by the AGN and stellar feedback depends on the total halo mass. Gas is easily blown out of smaller haloes, while it stays trapped in the deep potentials of galaxy clusters.

The different density profiles can be integrated to obtain the enclosed mass
\begin{equation}\label{massprofiles}
M_{\chi}(r)=4\pi\int_0^rds s^2\rho_{\chi}(s)\equiv f_{\chi}Y_{\chi}(r),
\end{equation}
where the subscript $\chi$ stands for both individual components and the total initial and final profiles ({\it dmo} and {\it bcm}, respectively). By construction the mass profiles $Y_{\chi}$ are normalised so that
\begin{equation}\label{massnorm}
M_{\rm nfw}(\infty)=Y_{\rm bgas}(\infty)=Y_{\rm egas}(\infty)=Y_{\rm cgal}(\infty)=Y_{\rm rdm}(\infty)\equiv M_{\rm tot}.
\end{equation}
Hence, the abundance of the components are solely governed by their fractions $f_{\chi}$.

The displacement function $d(r)$ can now be obtained by inverting the corrected mass profile ($M_{\rm bcm}$), i.e 
\begin{equation}
r_{\rm bcm}\equiv {\rm inv}\left[M_{\rm bcm}(r)\right],
\end{equation}
yielding
\begin{equation}\label{displfct}
d(r)=r_{\rm bcm}(M)-r,
\end{equation}
where $M\equiv 4\pi \Delta_{\rm 200}\rho_{c}r^3/3$. This defines a Lagrangian mapping from the initial ({\it dmo}) to the final ({\it bcm}) density profile and allows to directly shift particles in the simulation output as
\begin{equation}\label{rf}
r_f = r_i + d(r_i),
\end{equation}
in order to recover $\rho_{\rm bcm}$ starting from $\rho_{\rm dmo}$. Here $r_i$ and $r_f$ refer to the initial and final radii from the halo centre.

\begin{figure}[tbp]
\center{
\includegraphics[width=.325\textwidth,trim={0.0cm 0cm 0cm 0cm}]{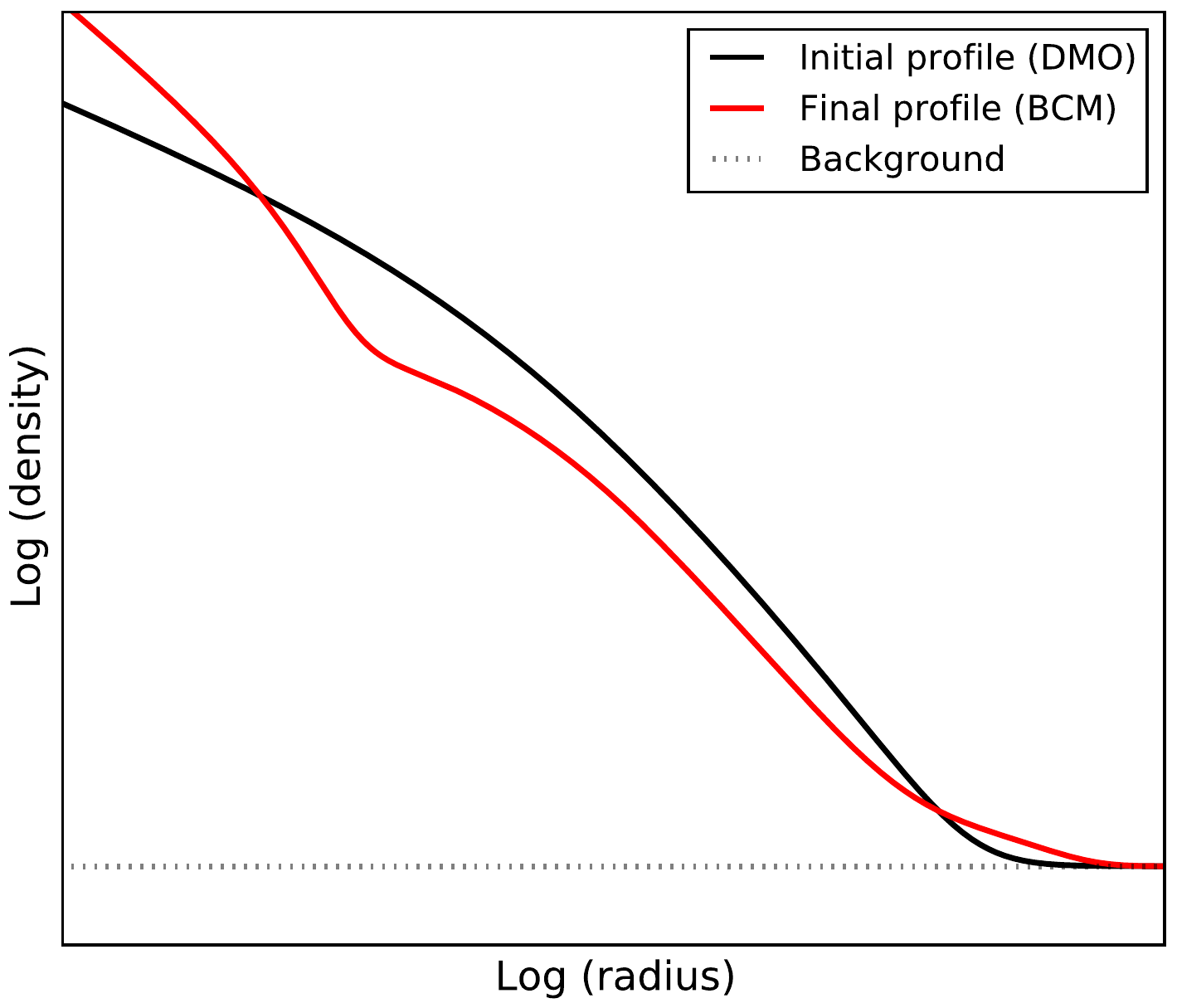}
\includegraphics[width=.325\textwidth,trim={0.0cm 0cm 0cm 0cm}]{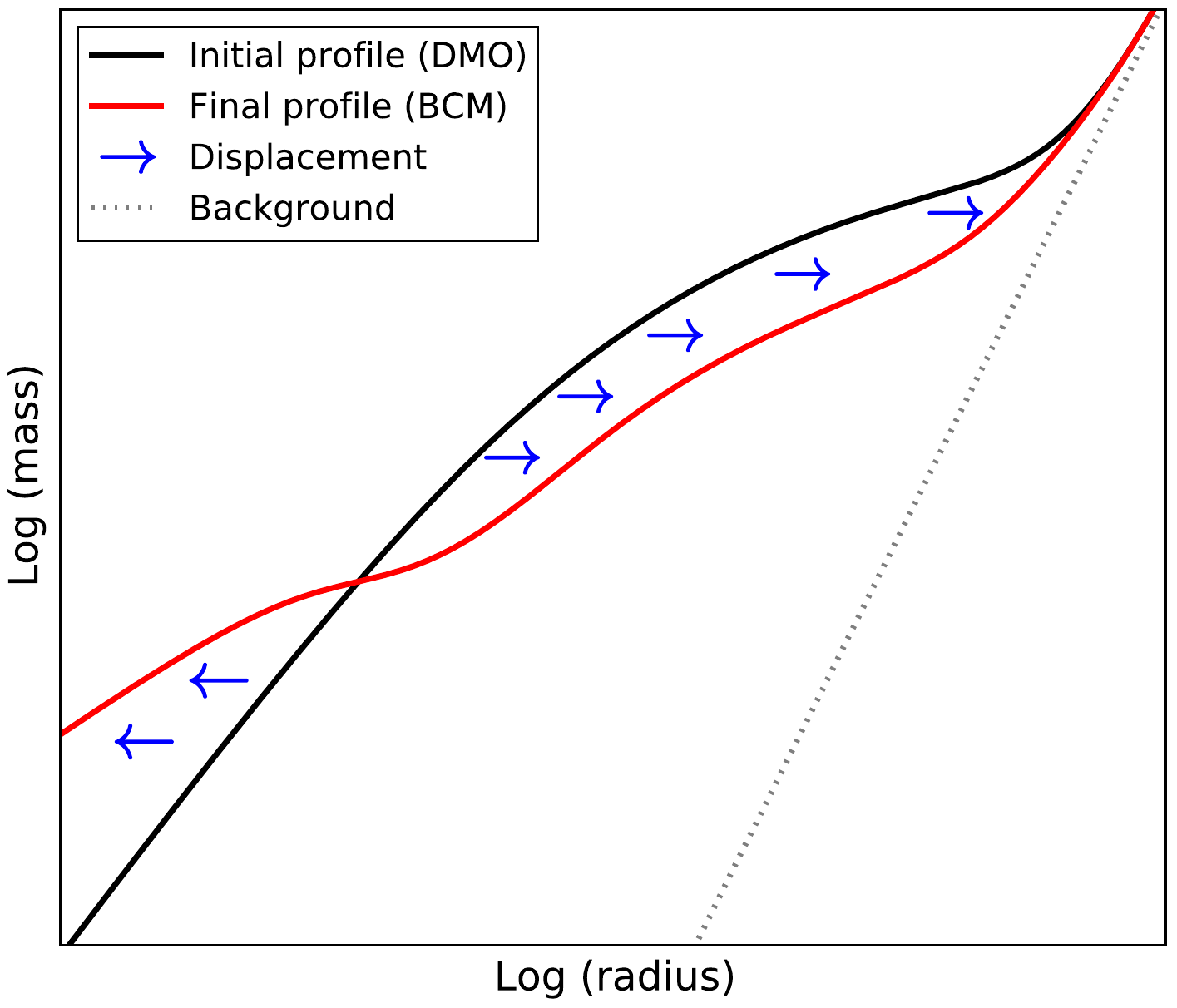}
\includegraphics[width=.325\textwidth,trim={0.0cm 0cm 0cm 0cm}]{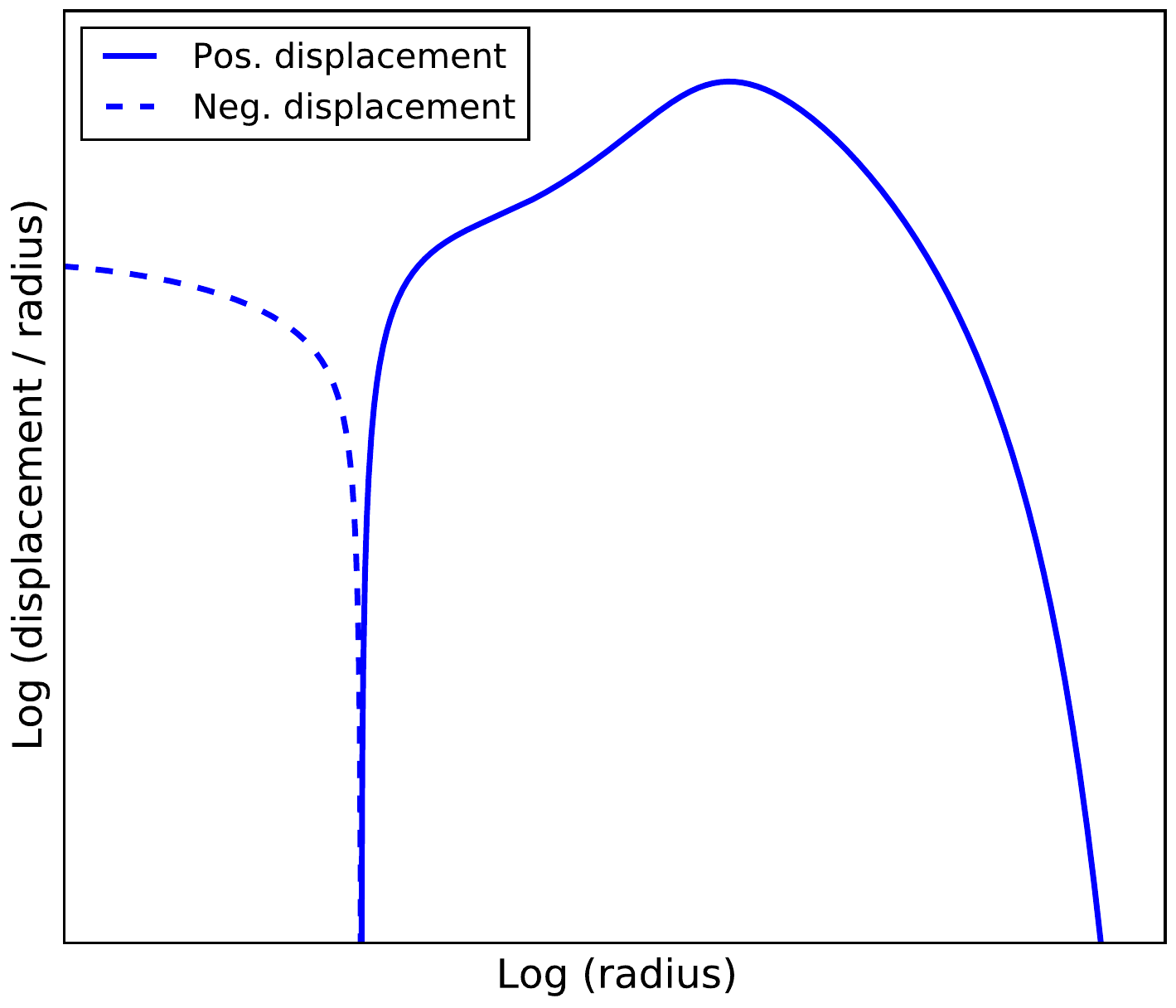}
\caption{\label{fig:displ}Schematic representation of the displacement model. {\it Left:} Initial (black) and final (red) density profiles with exaggerated baryonic fractions (for better illustration). {\it Middle:} Corresponding initial (black) and final (red) mass profiles and arrows indicating the displacement. {\it Right:} Relative displacement function (displacement didivided by radius). }}
\end{figure}

In Fig.~\ref{fig:displ} we illustrate the method by showing the density profile, mass profile, and displacement field of a schematic example with exaggerated baryonic effects. The left panel shows that the presence of baryons leads to a steepening of the inner and a flattening of the outer density profile, the former due to the central galaxy and the latter due to the bound and ejected gas components. The integrated mass profile (middle panel) is then used to define a displacement function (right panel) providing a Lagrangian mapping between the initial and final particle distribution around each halo. The displacement function is obtained by inverting the final mass profile and subtracting it from the reference radius as described in Eq. \eqref{displfct} and illustrated by the blue arrows in the middle panel.

\subsection{Initial profiles}\label{sec:rhoi}
This section contains the definitions of the DM-only profiles introduced in Eq.~\eqref{rhodmo} of the last section. We use a truncated version of the NFW-profile $\rho_{\rm nfw}$ \citep{Navarro1997} to avoid mass divergence. This is important for a proper normalisation of the different mass components, as shown in Eq.~\eqref{massnorm}. We follow Refs.~\citep{Baltz2009, Oguri2011} and use the profile
\begin{equation}\label{rhonfw}
\rho_{\rm nfw}(x,\tau)\equiv \frac{\rho_0}{x(1+x)^2}\frac{1}{(1+(x/\tau)^2)^2}\,,
\end{equation}
where $x=r/r_s$ and $\tau=r_{\rm tr}/r_{s}$ and where $r_s<r_{\rm 200}<r_{\rm tr}$ are the scale radius, virial radius, and truncation radius, respectively\footnote{Throughout the paper the viral mass and the viral radius are defined with respect to an overdensity criterion of 200 times the critical density, i.e $M_{200}=4\pi\Delta_{\rm 200}\rho_{c}r_{200}^3/3$ with $\Delta_{200}=200$.}. We tested the profile with haloes from DM-only simulations and obtain best results for $\tau=8 c$, where $c=r_{\rm 200}/r_{\rm s}$ is the concentration parameter. More about the truncated NFW profile as well as the integrated mass profile can be found in Appendix~\ref{app:NFW}.

The constant background component $\bar\rho_{bg}$ represents all non-collapsed matter. In a simulation of a volume $V$ and with the total amount of $N_p$ haloes, $\bar\rho_{bg}$ is given as
\begin{equation}\label{rhobg}
\bar\rho_{bg}=\Omega_{m}\rho_c - \frac{1}{V}\sum_i^{N_{\rm h}}M_{\rm tot}^{(i)},
\end{equation}
where the sum goes over all haloes in a simulation and $M_{\rm tot}^{(i)}$ is the total halo mass obtained by integrating over Eq.~\eqref{rhonfw}.

\subsection{Final profiles}\label{sec:rhof}
In this section we provide the definitions for the {\it bcm} profiles introduced in Eq.~\eqref{rhodmb} of Sec.~\ref{modelintro}. We start with the gas and stellar components and then discuss their back-reaction effect on the dark matter profile.

The bound gas in galaxy clusters mainly consists of X-ray emitting ionised hydrogen. Assuming hydrostatic equilibrium as well as a polytropic form for the gas pressure, i.e. $P=T\rho\propto\rho^{\Gamma}$, the bound gas profile can be described as \citep{Suto1998,Komatsu2001,Martizzi2013}
\begin{equation}\label{rhohga}
y_{\rm bgas}(x)=y_0\left[\frac{\ln(1+x)}{x}\right]^{\Gamma_{\rm eff}},\hspace{1cm}\Gamma_{\rm eff}=\frac{1}{\Gamma -1},
\end{equation}
where $x=r/r_s$. The polytropic index $\Gamma$ is fixed by assuming the slope of the gas profile to equal the one from the NFW profile at $r_{\rm 200}/\sqrt{5}$, yielding \citep{Komatsu2001}
\begin{equation}
\Gamma_{\rm eff}(c)=\frac{(1+3c/\sqrt{5})\ln(1+c/\sqrt{5})}{(1+c/\sqrt{5})\ln(1+c/\sqrt{5})-c/\sqrt{5}}\,.
\end{equation}
In our model the bound-gas profile is described by Eq.~\eqref{rhohga} until $r<r_{\rm 200}/\sqrt{5}$ and follows the NFW profile (i.e. Eq.~\ref{rhonfw}) further out. This is physically motivated as the gas is believed to be in hydrostatical equilibrium in the centre and to act like a collisionless fluid in the outskirts of the halo. Furthermore, the profile has been shown to agree well with both simulations \citep{Mohammed2014} and observational data \citep{Suto1998}.

The ejected gas profile is supposed to capture all the remaining gas which has neither been transformed into stars nor is part of the bound hot gas component, but has been expelled from the galaxy due to strong energy feedback from the AGN. The shape of the profile and especially how far it extends into inter-galactic space influences the power spectrum at rather large scales, a good estimate of the density profile is therefore important. We assume that the AGN energy induces velocity kicks to all gas particles following a Maxwell-Boltzmann distribution\footnote{This is motivated by the fact that prior to ejection the gas is believed to be dense enough so that Maxwell-Boltzmann statistics apply.}
\begin{equation}\label{MBdistr}
f_{\rm MB}(v)=\sqrt{\frac{2}{\pi}}\frac{v^2}{v_{\rm cr}^3}\exp\left[-\frac{v^2}{2v_{\rm cr}^2}\right].
\end{equation}
Furthermore, we assume the unbound gas to freely expand into space reaching the approximate distance $r=v t_0$ after a certain time $t_0$\footnote{This is clearly an over-simplification since gas particles slow down as they climb out of the halo potential. However, we will show later on that the effect can be fully absorbed by the free model parameter defined in Sec.~\ref{sec:rcr}.}. As a result, the Maxwell-Boltzmann distribution naturally transforms into the ejected-gas profile
\begin{equation}\label{rhoega}
y_{\rm egas}(r)=\frac{M_{\rm tot}}{4\pi r^2}f_{\rm MB}(r)=\frac{M_{\rm tot}}{(2\pi r_{\rm ej}^2)^{3/2}}\exp\left[-\frac{r^2}{2r_{\rm ej}^2}\right],
\end{equation}
where $r_{\rm ej}=v_{\rm cr}t_0$ determines how far the gas with critical velocity $v_{\rm cr}$ is pushed into inter-galactic space.

We will see later on that the choice of $r_{\rm ej}$ is crucial for studying the large-scale density distribution, as it determines up to what maximum scales baryonic effects are visible. In  Sec.~\ref{sec:rcr} two models for the halo mass dependence of $r_{\rm ej}$ are presented, one being heuristic and very simple and the other being physically motivated via the expected AGN energy input.

For the stellar profile of the central galaxy we follow Ref.~\citep{Mohammed2014} and use
\begin{equation}\label{rhocga}
y_{\rm cgal}(r)=\frac{M_{\rm tot}}{4\pi^{3/2}R_h}\frac{1}{r^2}\exp\left[-\left(\frac{r}{2R_h}\right)^2\right],\hspace{1cm}R_h=0.015\,r_{\rm 200},
\end{equation}
which behaves as a power law in the inner part and drops off exponentially well beyond the half-light radius $R_h$. The relation $R_h=0.015\,r_{\rm 200}$ is based on observations from Ref.~\citep{Kravtsov2014} and has been shown to agree with simulations of galaxy clusters  \citep{Mohammed2014}.

So far, we have defined specific profiles to model the baryonic fraction of the total matter distribution. There is, however, an additional dynamical modification of the DM component due to a back-reaction effect from the baryons which has to be taken into account. This effect is expected to contract the DM profile in the inner and expand it in the outer regions of the halo, the former because of additional mass from the central galaxy and the latter because of the missing gas in the halo outskirts.

The baryonic back-reaction effect can be included in the model by allowing for adiabatic relaxation (i.e. contraction and expansion) of the DM profile. Early models by the Refs. \citep{Barnes1984,Blumenthal1986} assumed shells of DM to contract due to the presence of the central galaxy while conserving angular momentum, i.e. $r_i M_i=r_fM_f$ ($M_f$ and $M_i$ being the total mass with and without central galaxy). More recent work by Refs.~\citep{Gnedin2004,Abadi2010} showed that a better agreement with numerical simulations is obtained if the slightly modified relation
\begin{equation}\label{ACmodel}
\frac{r_f}{r_i}-1=a\left(\frac{M_i}{M_f}-1\right),\hspace{1cm}a=0.68,
\end{equation}
is used instead. The authors of Refs.~\citep{Gnedin2004,Abadi2010} argue that reducing the constant $a$ to 0.68 (with respect to $a=1$ for angular momentum conservation) corrects for the fact that the growth of the central galaxy is not an instantaneous process. In Ref.~\citep{Teyssier2011} it is furthermore shown that Eq.~\eqref{ACmodel} is not only a good model for the adiabatic contraction induced by the central galaxy but can also account for the back-reaction due to missing gas in the halo outskirts. We follow Ref.~\citep{Teyssier2011} and write for the mass terms
\begin{equation}
\begin{split}
M_i&=M_{\rm nfw}(r_i),\\
M_f&=f_{\rm cdm}M_{\rm nfw}(r_i)+f_{\rm cgal}Y_{\rm cgal}(r_f)+f_{\rm bgas}Y_{\rm bgas}(r_f) + f_{\rm egas}Y_{\rm egas}(r_f),
\end{split}
\end{equation}
before solving iteratively for $\xi = r_f/r_i$ in Eq.~\eqref{ACmodel}. Once $\xi$ is known, the DM mass profile with baryonic back-reaction is simply $M_{\rm rdm}(r)=M_{\rm nfw}(r/\xi)$ yielding the density profile
\begin{equation}
y_{\rm rdm} (r)=\frac{1}{4\pi r^2}\frac{d}{dr}M_{\rm nfw}(r/\xi).
\end{equation}

\subsection{Abundance fractions}\label{sec:frac}
By construction, the sum over all component fractions equals one, i.e.
\begin{equation}
f_{\rm rdm} + f_{\rm bgas}(M)+f_{\rm egas}(M)+f_{\rm cgal}(M)=1.
\end{equation}
While the relaxed DM fraction is a constant given by $f_{\rm rdm}=1-\Omega_b/\Omega_m$ (where $\Omega_b/\Omega_m$ is the cosmic baryon fraction), the stellar fraction and the gas fractions depend on the total mass of the halo. For the bound and ejected gas components the mass dependence comes from energy injection of AGN feedback. While in large clusters the gas may only be heated up by the AGN activity, it is ejected outside of the halo in the case of Milky-Way sized galaxies.

The hot gas inside galaxy clusters can be observed via X-ray radiation. The signal mainly comes from within $r_{500}$ and is therefore probing the bound gas component ({\it bgas}). Based on X-ray observations, the bound gas fraction can be parametrised as follows
\begin{equation}\label{fhga}
f_{\rm bgas}(M)=\frac{\Omega_b/\Omega_m}{1+(M_c/M)^{\beta}},
\end{equation}
where $M_c$ and $\beta$ are free model parameters describing the mass scale and the slope of the hot-gas suppression towards small halo masses \citep{Mohammed2014}.

\begin{figure}[tbp]
\center{
\includegraphics[width=.47\textwidth,trim={1cm 1cm 3cm 1cm}]{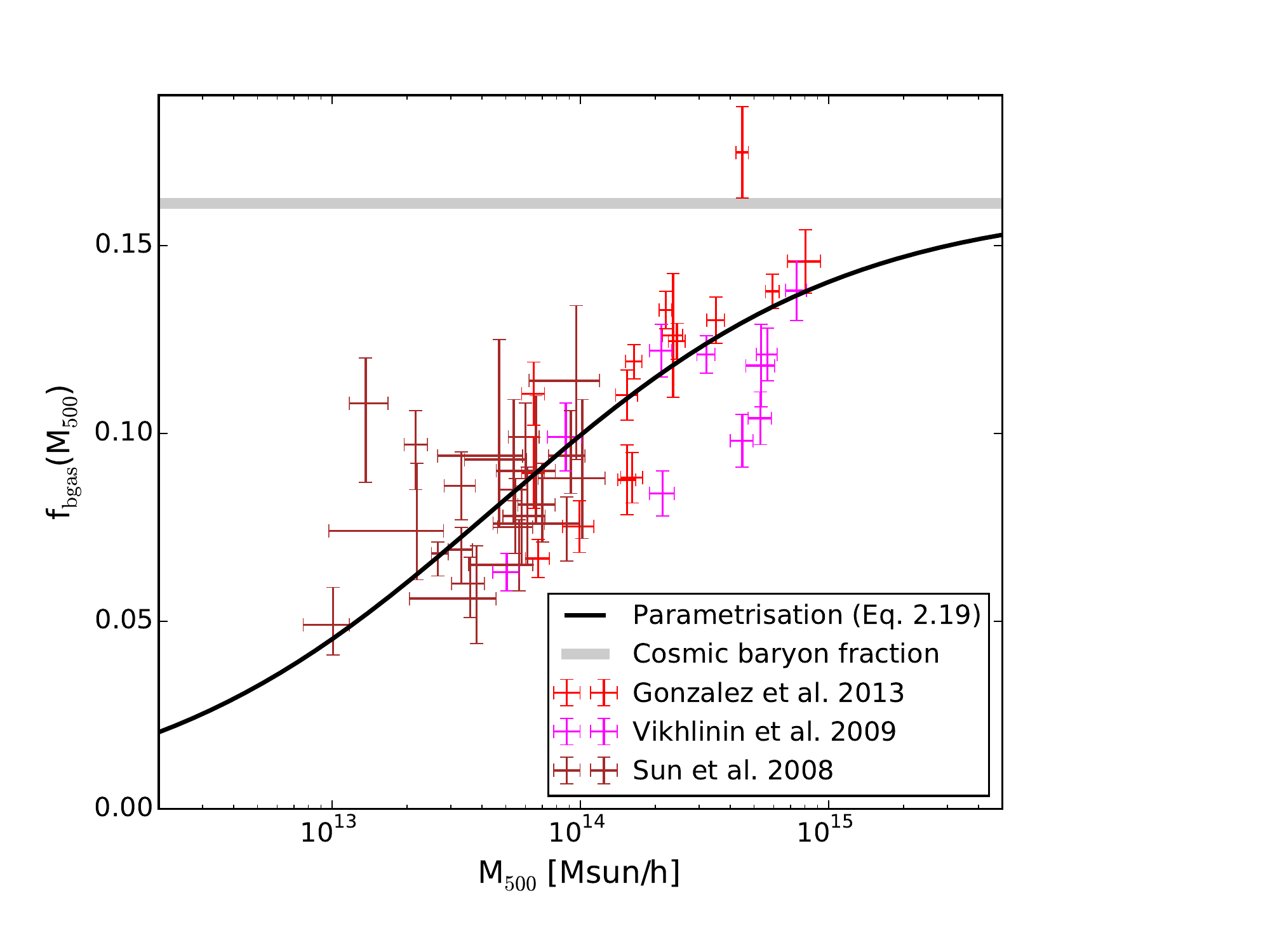}
\includegraphics[width=.47\textwidth,trim={1cm 1cm 3cm 1cm}]{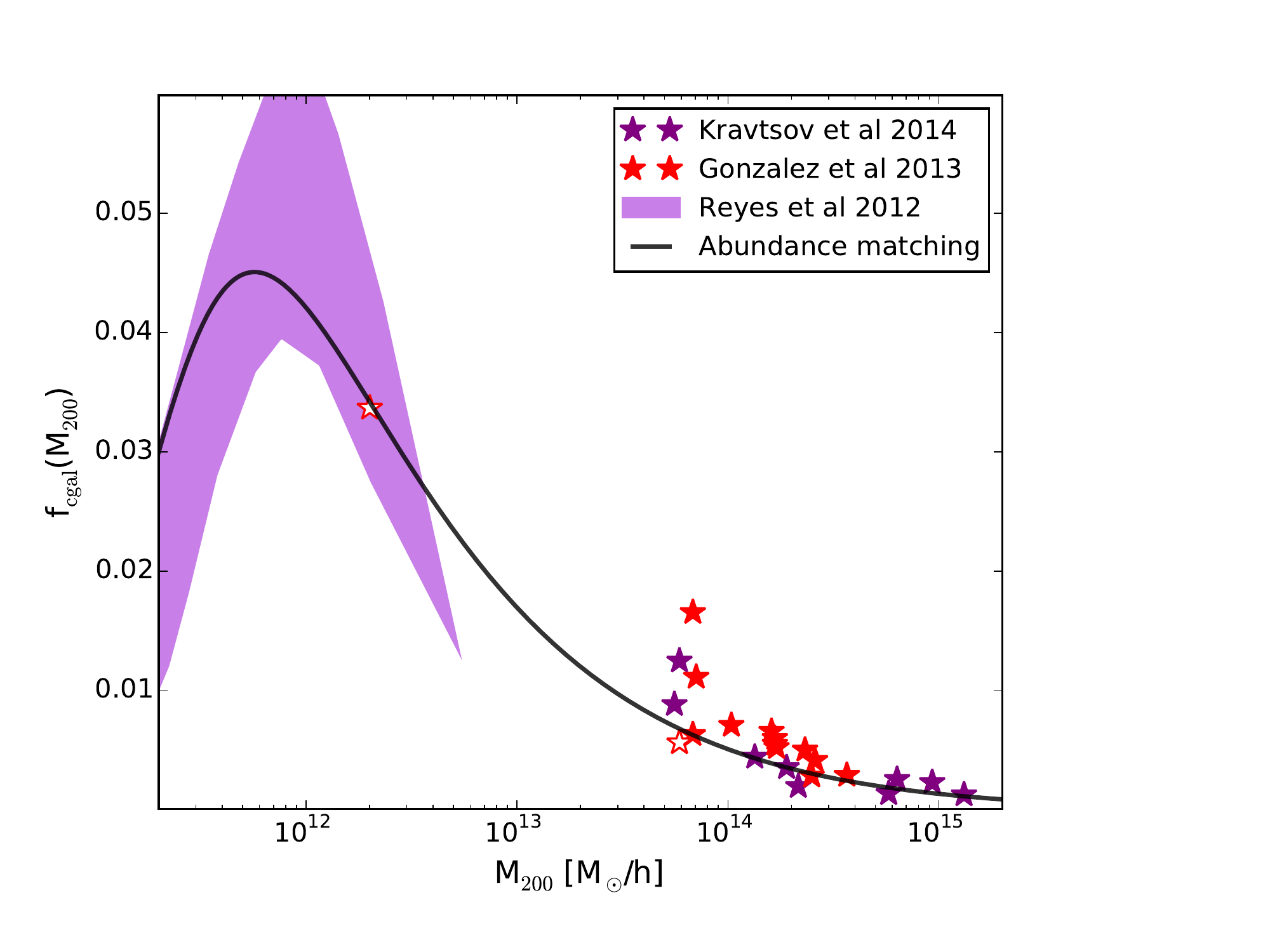}
\caption{\label{fig:fhga}Left panel: Observed fraction of X-ray emitting bound gas (red, pink, and brown symbols) compared to the two-parameter fit of Eq.~\eqref{fhga} with best fitting parameters $\beta$=0.6 and M$_c$=1.2e14 M$_{\odot}$/h (black line). Right panel: Observed stellar fraction of the central galaxy with mass estimates from X-ray observations (stars) and weak lensing (pink area) compared to the parametrisation from Eq.~\eqref{fcga} based on abundance matching \citep{Kravtsov2014} (black line).}}
\end{figure}

In the left-hand-side panel of Fig.~\ref{fig:fhga} we plot observational X-ray data from the Refs.~\citep{Gonzalez2013,Vikhlinin2009,Sun2008} together with the fit from Eq.~\eqref{fhga}. The bulk of the observations lie below the baryon fraction given by Planck (horizontal dotted line), showing that a significant fraction of the gas has been pushed out beyond $r_{500}$. Furthermore, the clear trend in the data suggests that smaller haloes loose more of their gas due feedback effects. The trend is well described by Eq.~\eqref{fhga} with the best fitting parameters $M_c=1.2\times10^{14}$ M$_{\odot}$/h and $\beta=0.6$. However, as the data is subject to significant statistical and potentially systematical errors, we probe wide ranges of values for $M_c$ and $\beta$ as well as the possibility of scatter in this paper.

The fraction of stars in the central galaxy can be parametrised assuming abundance matching. Ref.~\citep{Kravtsov2014} proposed a functional form
\begin{equation}\label{fcga}
\begin{split}
f_{\rm cgal}(M)&=\epsilon \left(\frac{M_1}{M}\right) 10^{g(\log_{10}[M/M_1])-g(0)}\\
g(x)&=-\log_{10}[10^{\alpha x}+1] + \delta\frac{\left(\log_{10}[1+\exp(x)]\right)^{\gamma}}{1+\exp(10^x)}
\end{split}
\end{equation}
based on~\citep{Behroozi2013} with updated parameters $\epsilon=0.023$, $M_1=1.526\times10^{11}$ M$_\odot$/h, $\alpha=-1.779$, $\delta=4.394$ and $\gamma=0.547$.

In the right-hand-side panel of Fig.~\ref{fig:fhga} we plot the stellar fraction of the central galaxy from X-ray \citep{Gonzalez2013,Kravtsov2014} and weak lensing data \citep{Reyes2012} given by star symbols and the pink shaded area. The abundance-matching curve from Eq.~\eqref{fcga} is added as black solid line and agrees well with the direct observations as shown by Ref.~\citep{Kravtsov2014}.

Based on the observations of the bound gas and stars, it is now straight forward to estimate the ejected gas fraction by subtracting the former components from the total baryon abundance, i.e.,
\begin{equation}\label{fega}
f_{\rm egas}(M)=\Omega_b/\Omega_m-f_{\rm bgas}(M)-f_{\rm cgal}(M).
\end{equation}
The ejected gas is therefore assumed to make up for the remaining unobserved baryons expected from the {\it Planck} CMB measurement  \citep{Ade2014}.

\subsection{Ejected gas radius}\label{sec:rcr}
We model the ejected gas component with a Gaussian profile defined in Sec.~\ref{sec:rhof}. This is motivated by assuming that the velocity-kicks from the AGN follow a Maxwell-Boltzmann distribution and escape the haloes with constant speed. What we have not done so far is to estimate the critical ejected gas radius ($r_{\rm ej}$) defined in Eq.~\eqref{rhoega}. Later on in Sec.~\ref{sec:ps}, we will show that the value of $r_{\rm ej}$ is important as it sets the minimum wave mode above which the power spectrum is suppressed by baryons.

A natural assumption is to link the critical radius to the typical distance travelled by a gas particle with the halo escape velocity $v_{\rm esc}$. The escape radius is given by
\begin{equation}\label{resc}
r_{\rm esc}\equiv t_0 v_{\rm esc}\sim t_0\sqrt{\frac{8\pi}{3} G\Delta_{\rm 200}\rho_c}\, r_{\rm 200}\sim 0.5\sqrt{\Delta_{\rm 200}}\,r_{\rm 200},
\end{equation}
where we approximate the typical time-scale $t_0$ by half the Hubble time. The most simple physically motivated model is to assume
\begin{equation}\label{rcr1}
r_{\rm ej}\equiv\eta_{\rm a} r_{\rm esc}
\end{equation}
with the free model parameter $\eta_a$, where the ejection radius is simply proportional to the virial radius. We will call this heuristic approach model (A) throughout the paper. A more evolved model can be constructed by using the efficiency of gas ejection given by the observationally constrained expelled gas fraction ($f_{\rm egas}$). The value of $f_{\rm egas}$ determines how many gas particles obtain velocity-kicks beyond the escape velocity, which sets the width of the Maxwell-Boltzmann distribution. Integrating Eq.~\eqref{MBdistr} and equating it with the gas fraction yields\begin{equation}\label{rcr2}
1.0 - {\rm Erf}\left[\frac{\eta_{\rm b}r_{\rm esc}}{\sqrt{2}r_{\rm ej}}\right] + \sqrt{\frac{2}{\pi}}\frac{\eta_{\rm b}r_{\rm esc}}{r_{\rm ej}}\exp\left[-\frac{\eta_{\rm b}^2r_{\rm esc}^2}{2 r_{\rm ej}^2}\right]\equiv \frac{\Omega_m}{\Omega_b} f_{\rm egas}(M),
\end{equation}
which can be solved iteratively to determine $r_{\rm ej}$. This consists of a more evolved approach referred to as model (B) in this paper. It also has one free parameter $\eta_{\rm b}$ to account for uncertainties in the approximative calculation of $r_{\rm esc}$.

The free model parameter $\eta_\chi$ (with $\chi=a,b$ depending on the model) does not only incorporate uncertainties related to the time-span  $t_0$ of AGN activity, but it also accounts for the fact that gas does not escape with constant speed (as assumed in Sec.~\ref{sec:rhof}). The velocity decrease of gas escaping a spherical NFW potential is indeed proportional to the viral radius and can therefore be absorbed into the $\eta_\chi$-parameter.

\begin{figure}[tbp]
\center{
\includegraphics[width=0.9\textwidth,trim={1.5cm 6.5cm 4.5cm 2cm}]{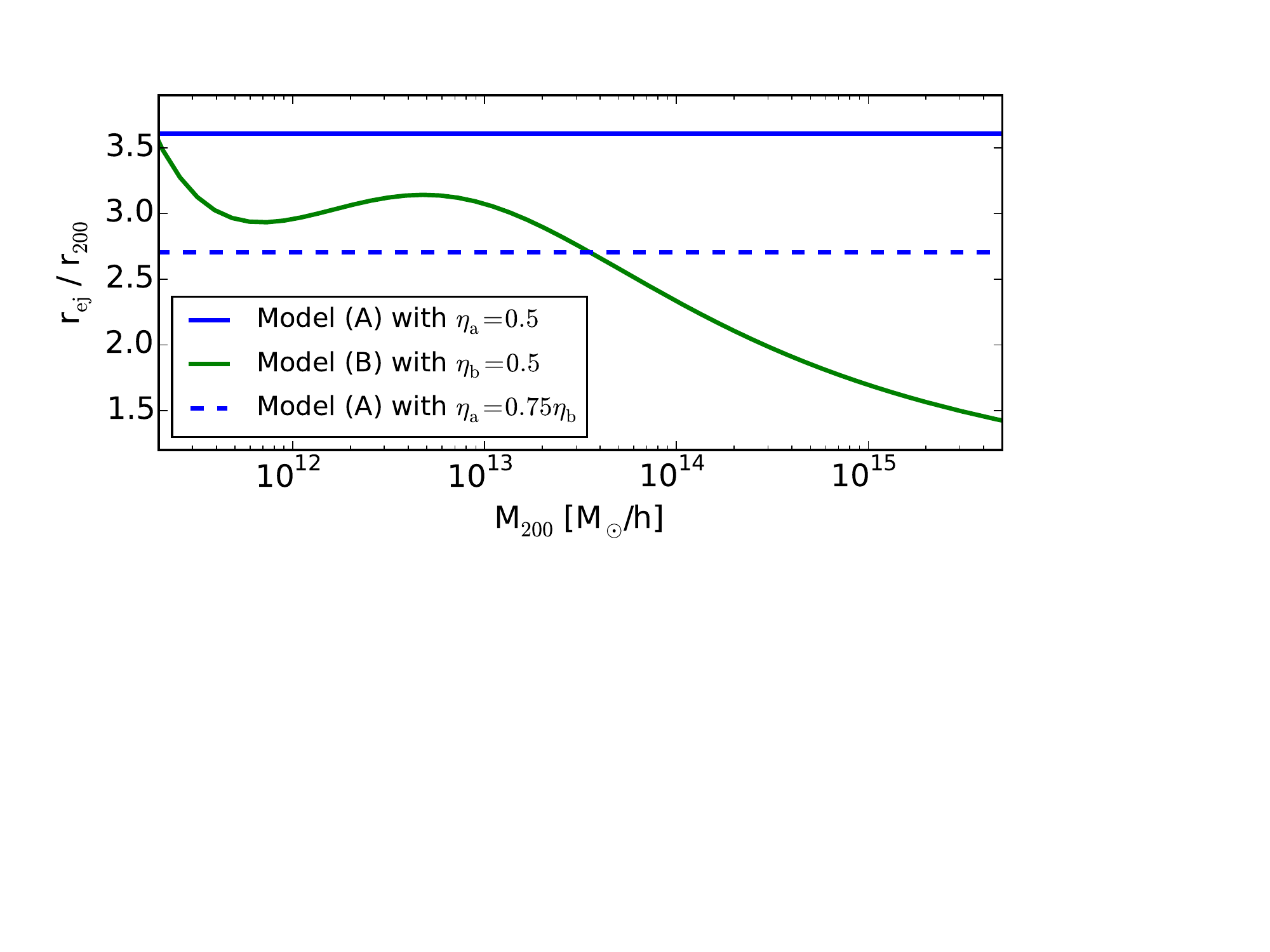}
\caption{\label{fig:rcr}Critical radius of ejected gas in model (A) and model (B).}}
\end{figure}

In Fig.~\ref{fig:rcr} the two models are compared to each other, the blue and green lines showing the critical radius obtained from model (A) and (B). We have investigated how the power spectrum is affected by using either model (A) or model (B) and it turns out that both models yield nearly identical results for the parameter values
\begin{equation}
\eta_{\rm a}=0.75\,\eta_{\rm b}.
\end{equation}
This is intuitively understandable by looking at the dashed blue line in Fig.~\ref{fig:rcr} which lines up closely with the green solid line of model (B), especially in the regime  $M=10^{12}$ - $10^{15}$ M$_{\odot}$/h most relevant for the power spectrum measurement.

\section{\boldmath Correcting $N$-body simulations}\label{sec:sims}
The {\it baryonic correction model} presented in the last section can be applied to outputs of $N$-body simulations. The displacement function defined by Eq.~\eqref{displfct} allows to shift particles around haloes, thereby modifying the initial NFW profiles to account for the effects of baryons. In this section we present our simulations and we show how the matter distribution is modified by the particle displacement.

\subsection{Setup of simulations}
We use $N$-body simulations from previous work \citep{Schneider2015,Schneider2015b} (performed with {\tt Pkdgrav3} \citep{Stadel2001}) with box sizes of $L=$ 32, 64 128, 256 Mpc/h and particle numbers of $N=$ 64 128, 256, 512 per dimension, resulting in a fixed particle mass resolution of $M_p=10^{10}$ M$_\odot$/h\footnote{This has been shown to be sufficient for power spectrum measurements \citep{vanDaalen2015,Schneider2015}.}. The initial conditions are set up at redshift 50 using second order Lagrangian perturbation theory (2LPT). For the cosmological parameters we choose $\Omega_m=0.3071$, $\Omega_{\Lambda}=0.6929$, $\Omega_b=0.0483$, $h = 0.6777$, $n_s = 0.9611$, and $\sigma_8 = 0.8288$ obtained by the {\it Planck} mission \citep{Ade2014}.

The halo positions and concentrations are determined with the help of the {\tt AHF} halo finder \citep{Gill2004,Knollmann2009}, assuming an overdensity criterion of $\Delta_{\rm vir}=200$ times the critical density $\rho_c$. 
In principle, the model is independent of the halo finding technique and also works for other common definitions of the halo mass. However, it crucially relies on halo concentrations and therefore requires an accurate measurement of the halo profiles. Haloes with fewer than 500 particles are not included in the analysis, as their concentration cannot be measured accurately. We have verified that this choice does not affect our final results significantly.

\begin{figure}[tbp]
\center{
(a)\hspace{4.75cm} (b)\hspace{4.75cm} (c)
\vspace{0.3cm}

\includegraphics[width=.323\textwidth,trim={0cm 0.0cm -0.2cm 0cm}]{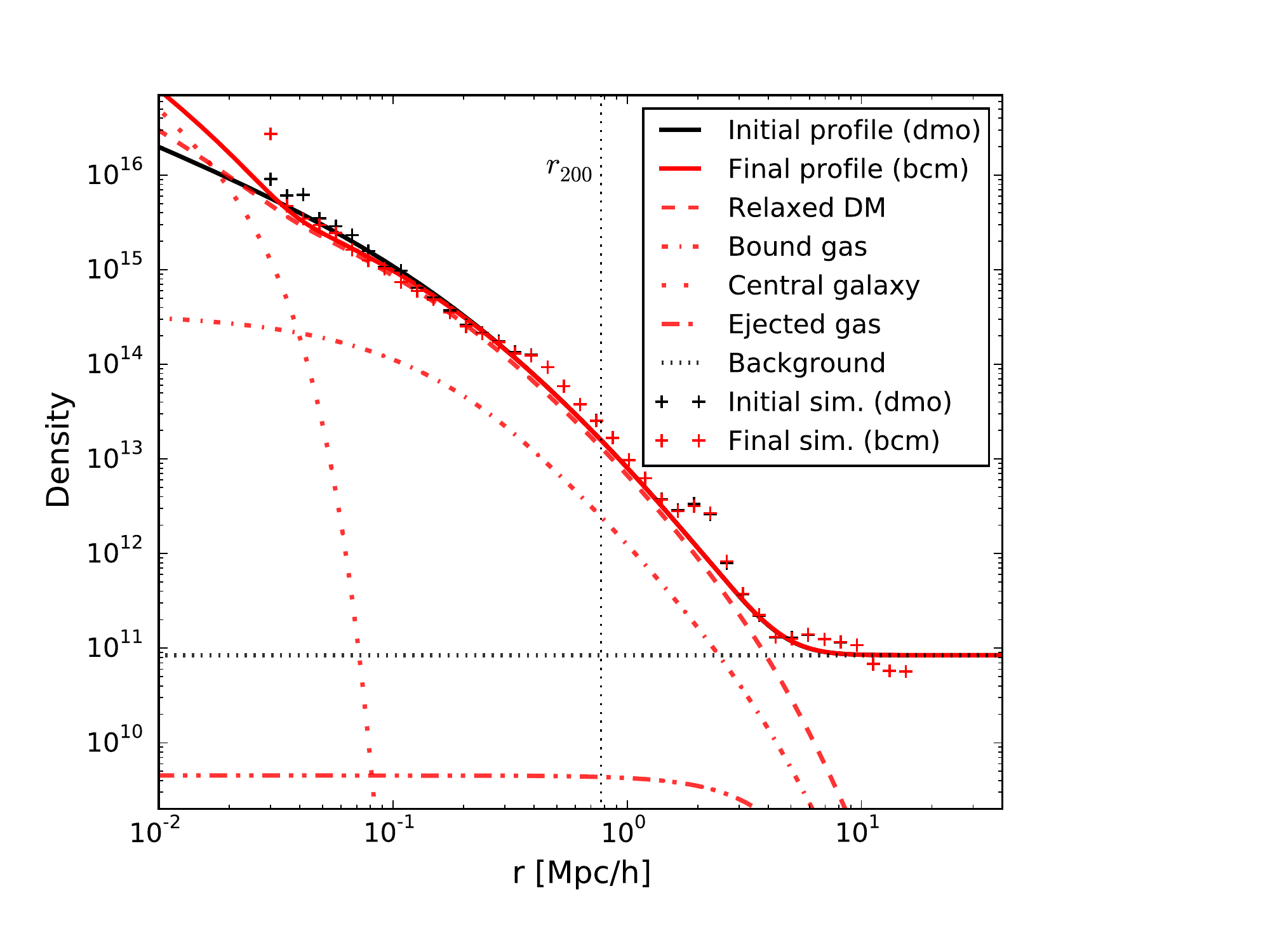}
\includegraphics[width=.323\textwidth,trim={0cm 0.0cm -0.2cm 0cm}]{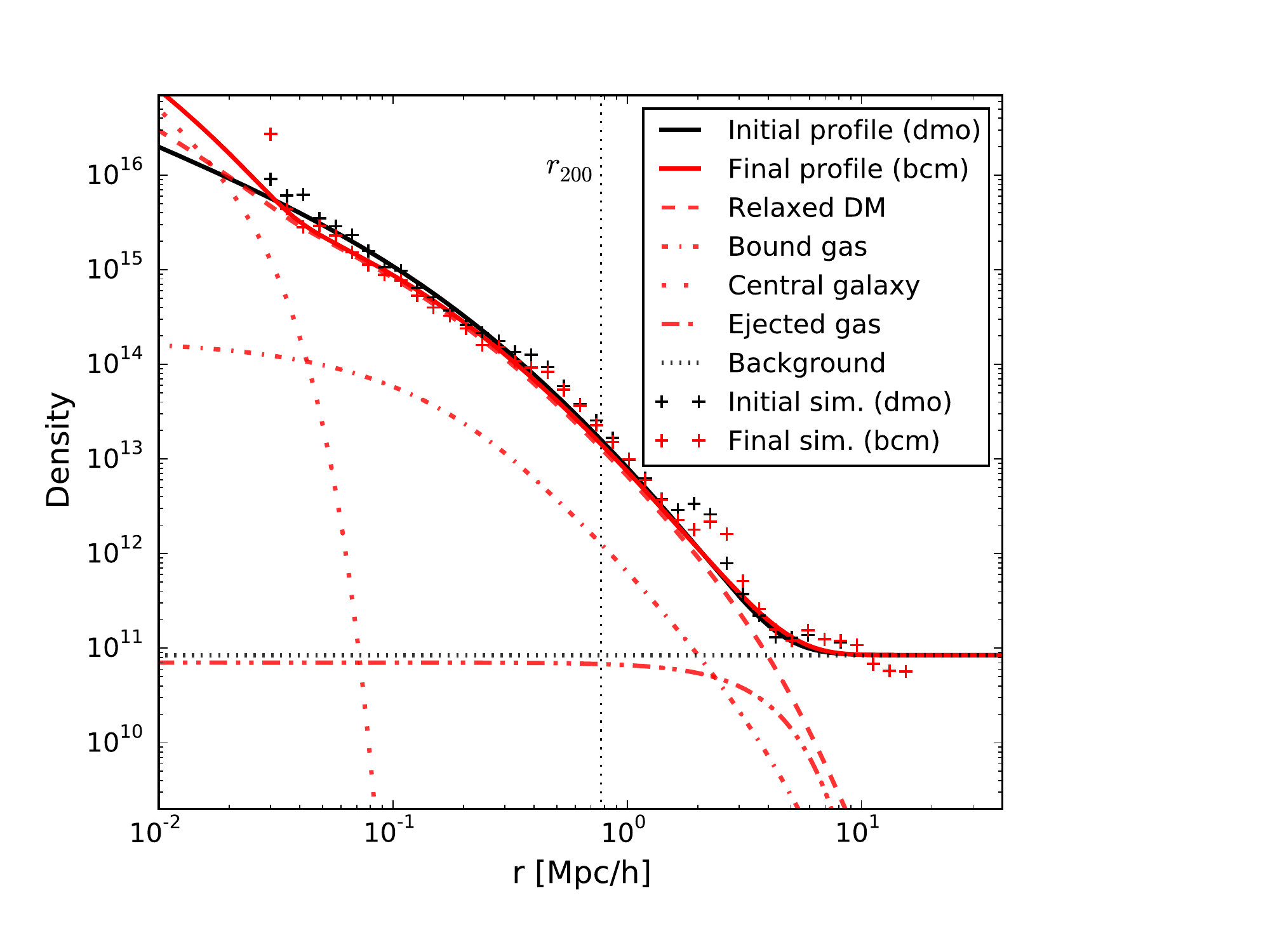}
\includegraphics[width=.323\textwidth,trim={0cm 0.0cm -0.2cm 0cm}]{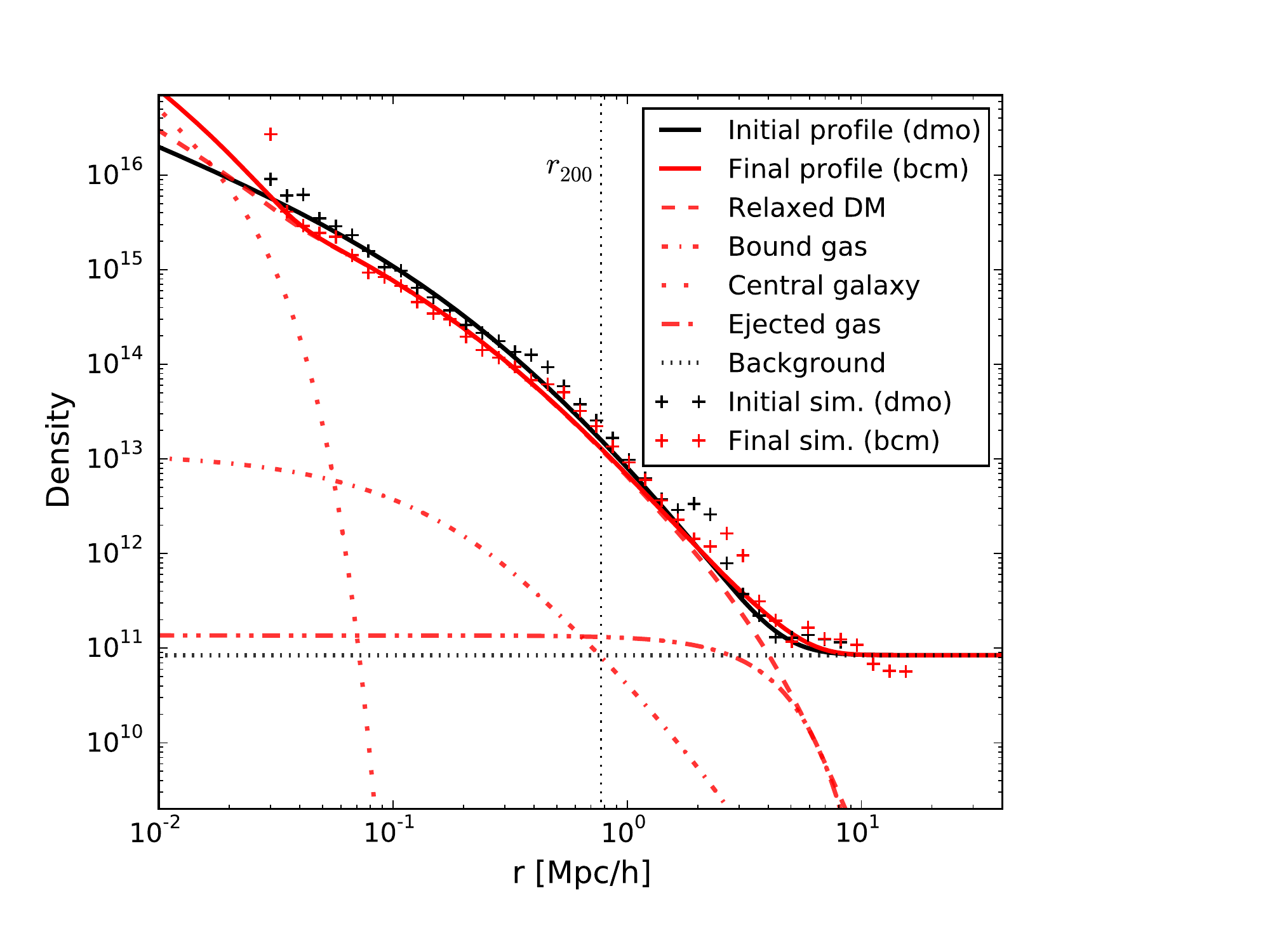}\\
\includegraphics[width=.323\textwidth,trim={0cm 0.0cm -0.2cm 0cm}]{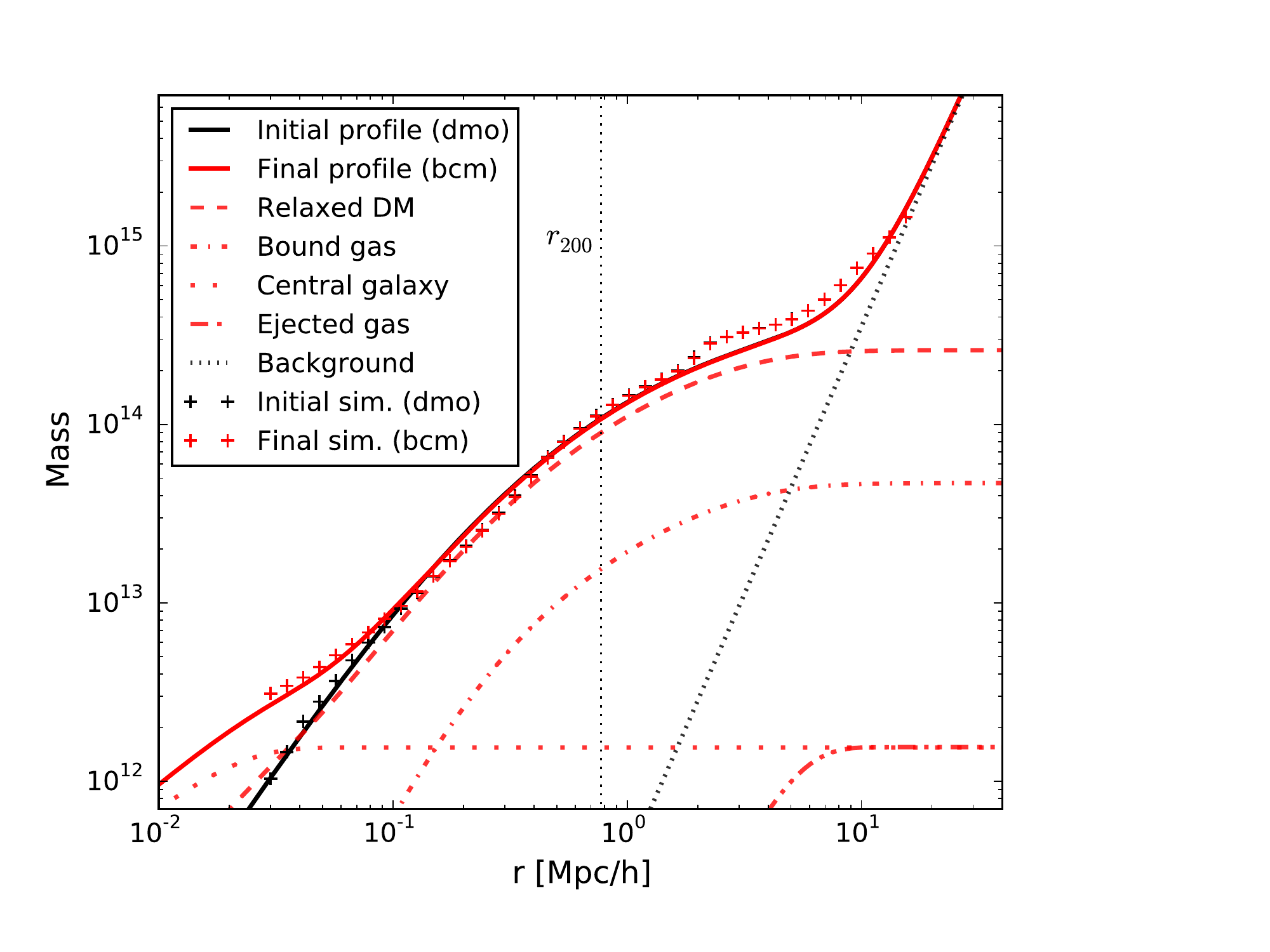}
\includegraphics[width=.323\textwidth,trim={0cm 0.0cm -0.2cm 0cm}]{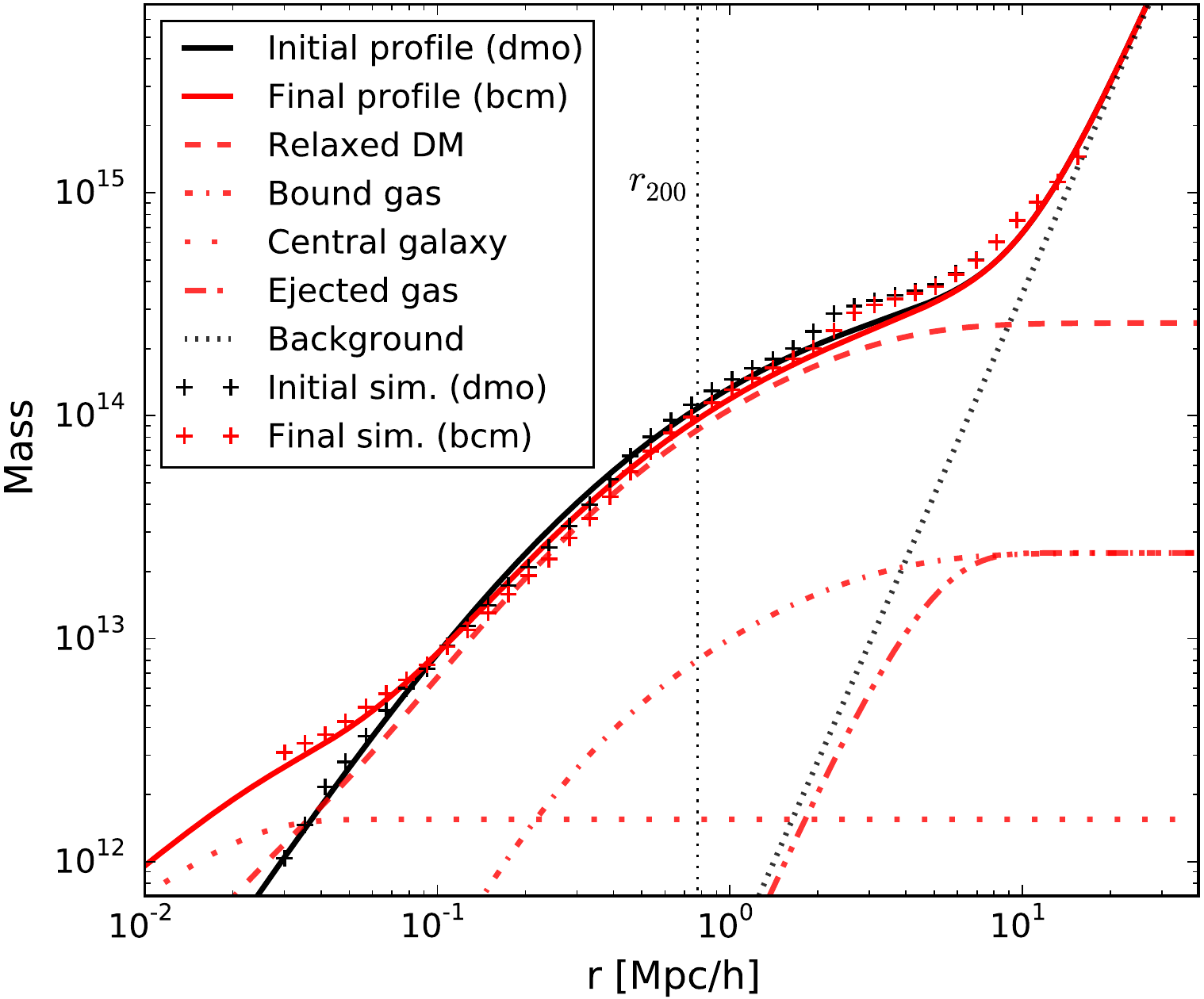}
\includegraphics[width=.323\textwidth,trim={0cm 0.0cm -0.2cm 0cm}]{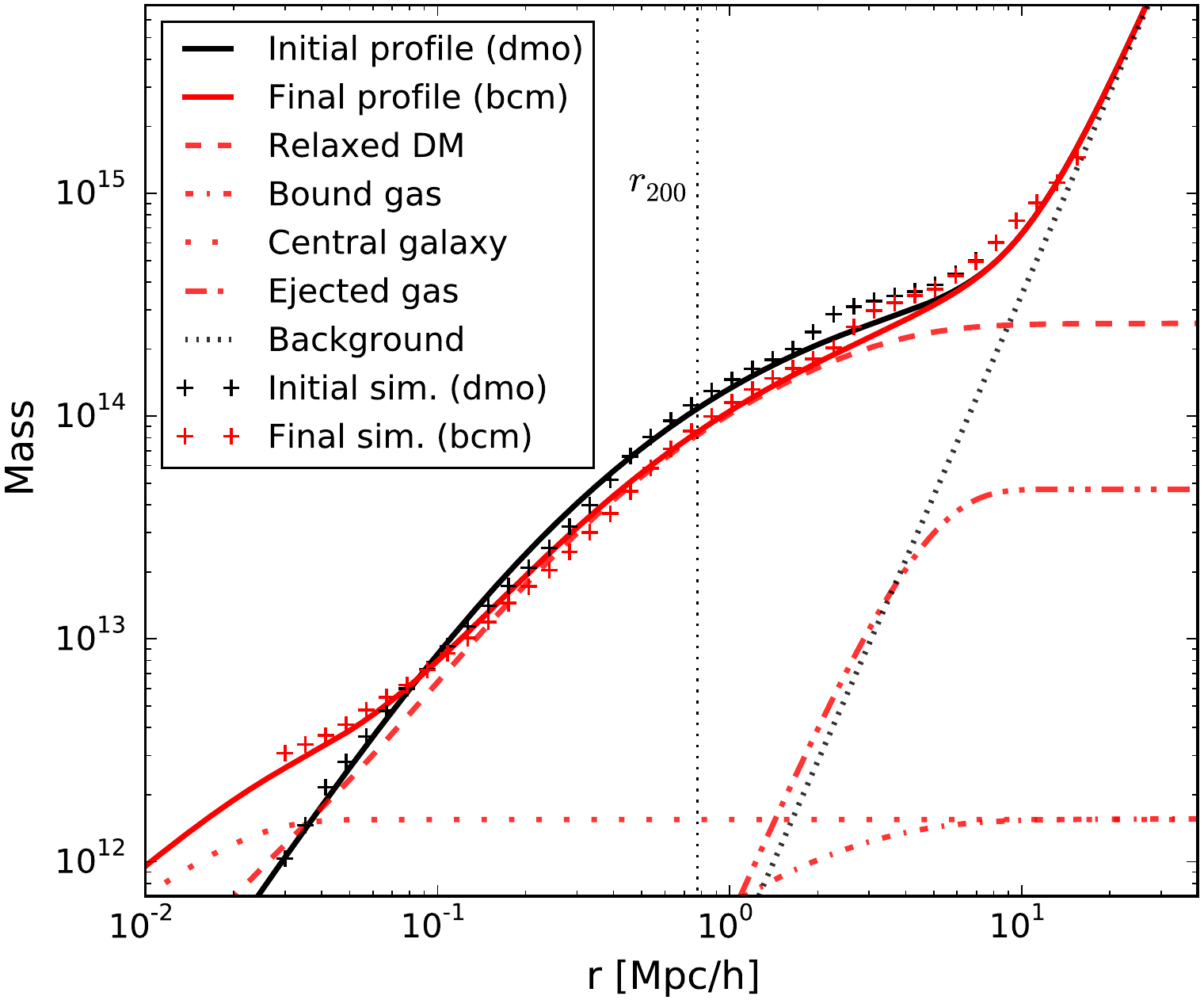}\\
\includegraphics[width=.323\textwidth,trim={0cm 0.0cm -0.2cm 0cm}]{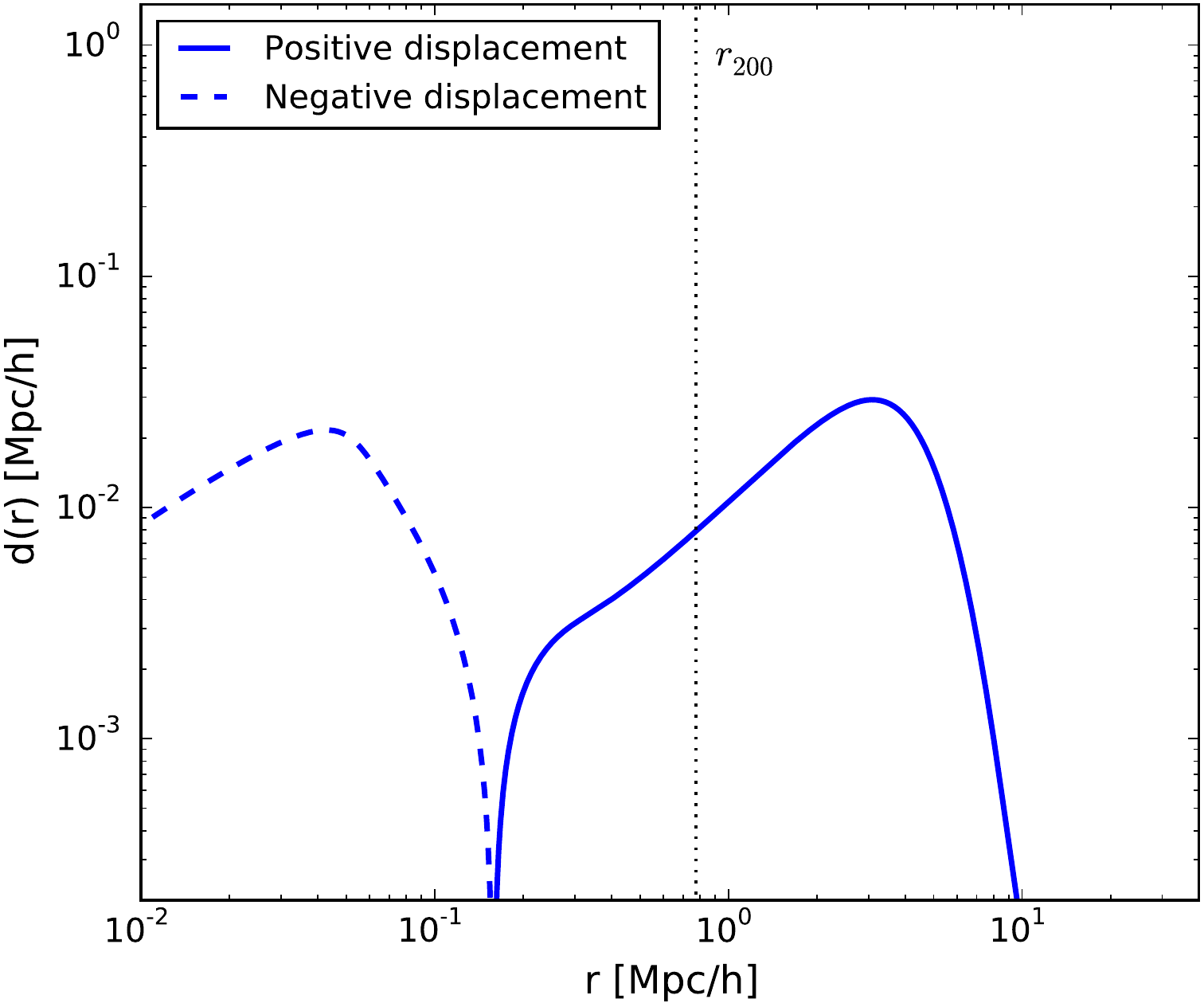}
\includegraphics[width=.323\textwidth,trim={0cm 0.0cm -0.2cm 0cm}]{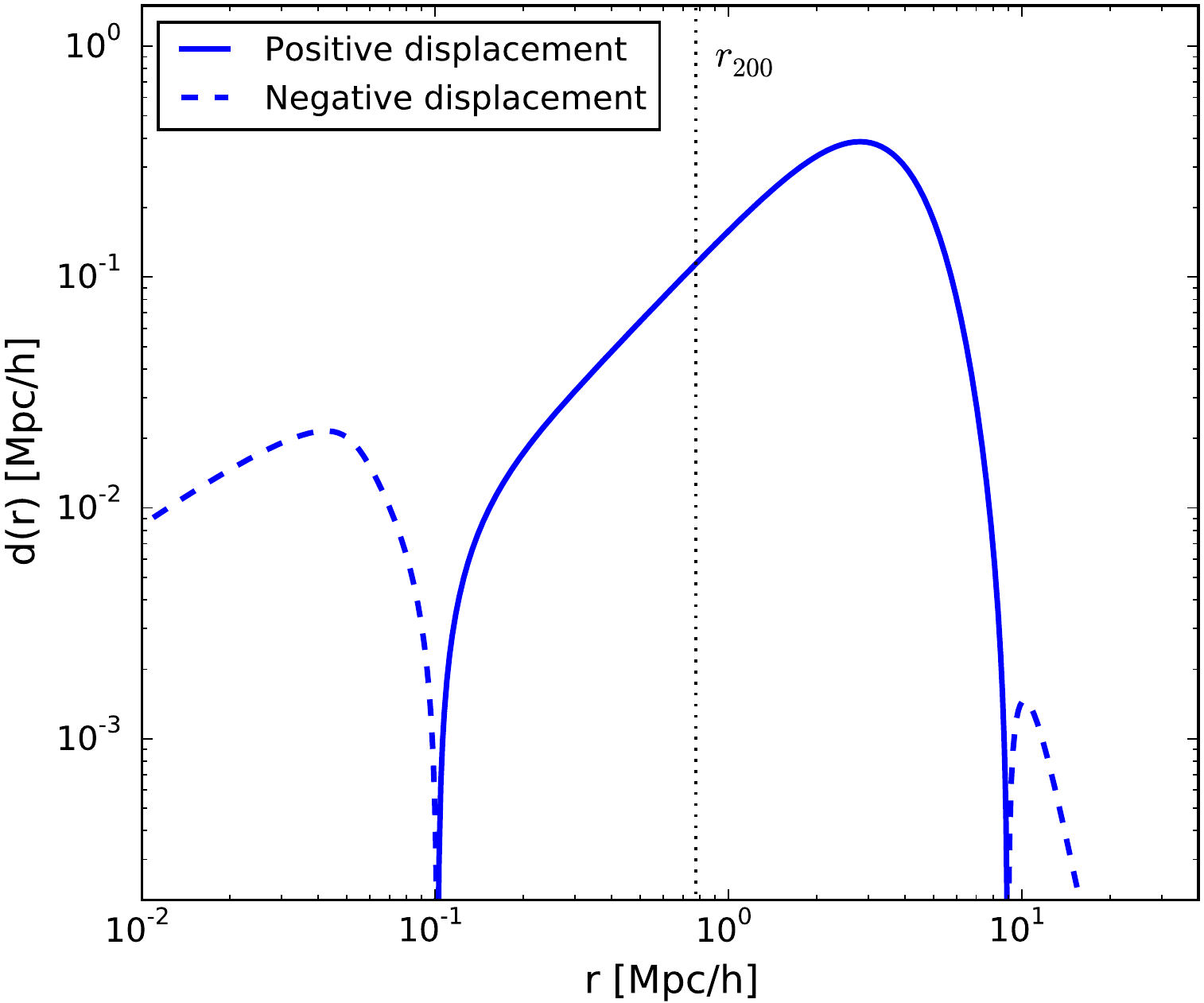}
\includegraphics[width=.323\textwidth,trim={0cm 0.0cm -0.2cm 0cm}]{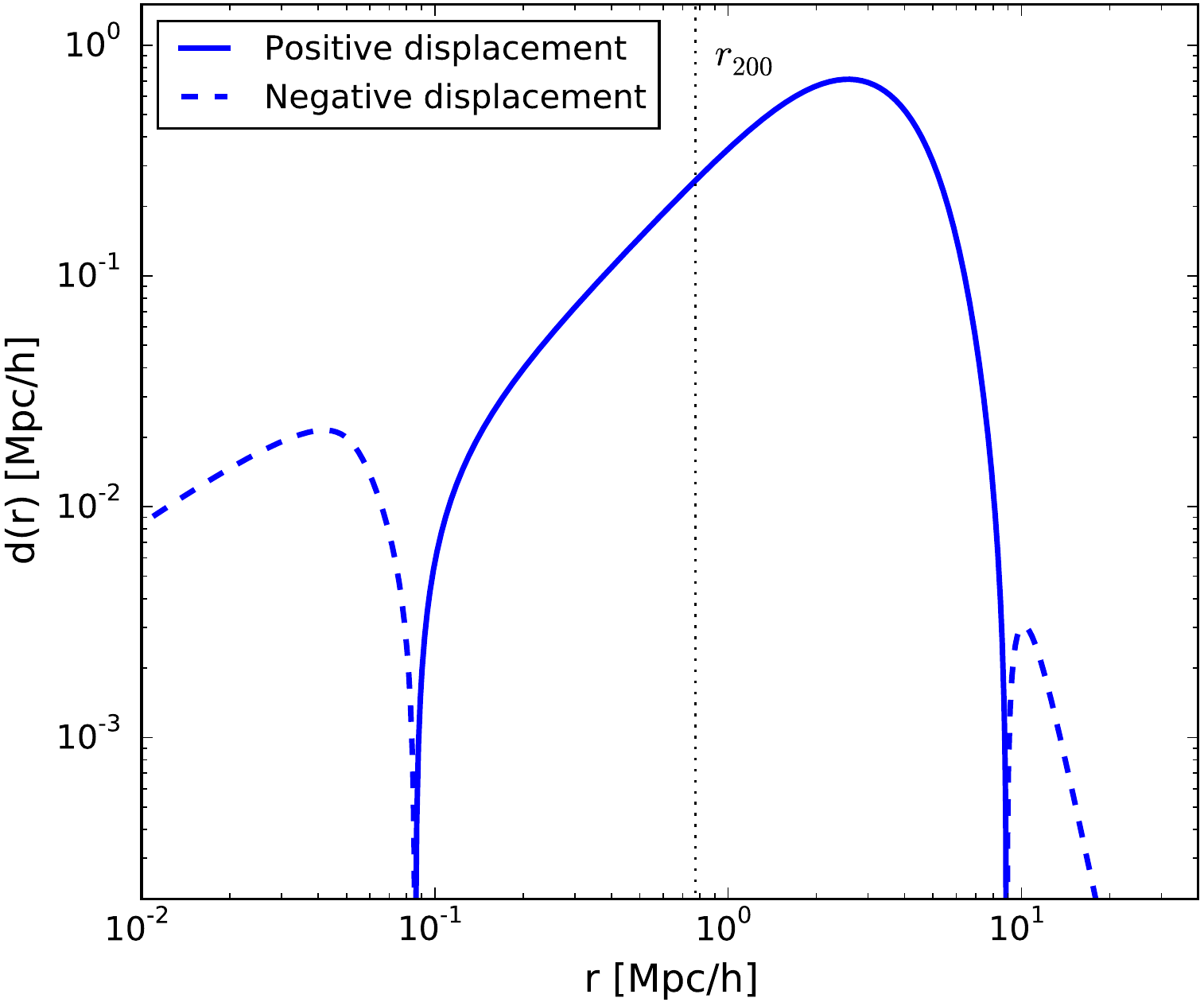}\\
\hspace{0.28cm}\includegraphics[width=.312\textwidth,trim={-1.cm -0cm -1.cm -0.4cm}]{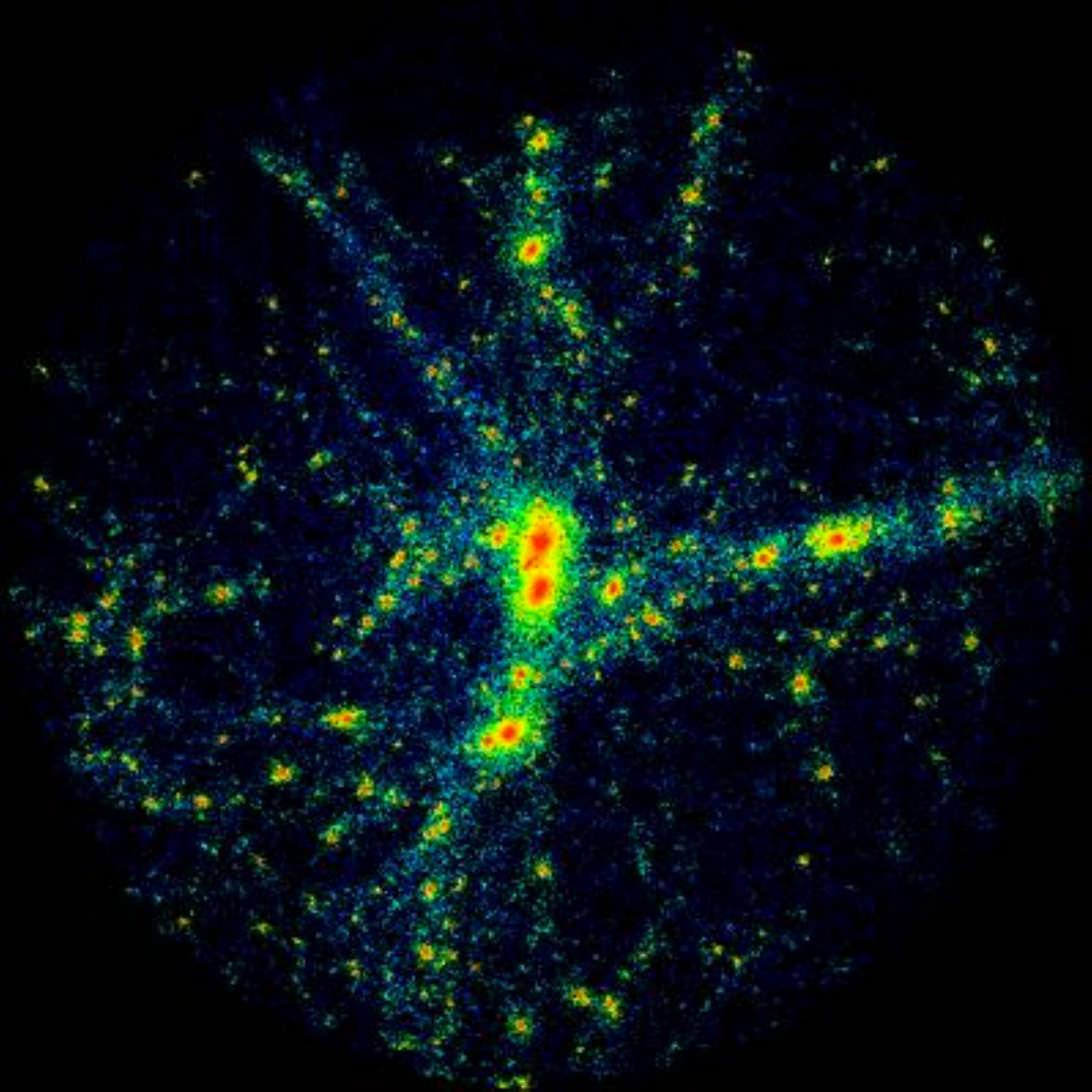}
\hspace{0.24cm}\includegraphics[width=.312\textwidth,trim={-1.cm -0cm -1.cm -0.4cm}]{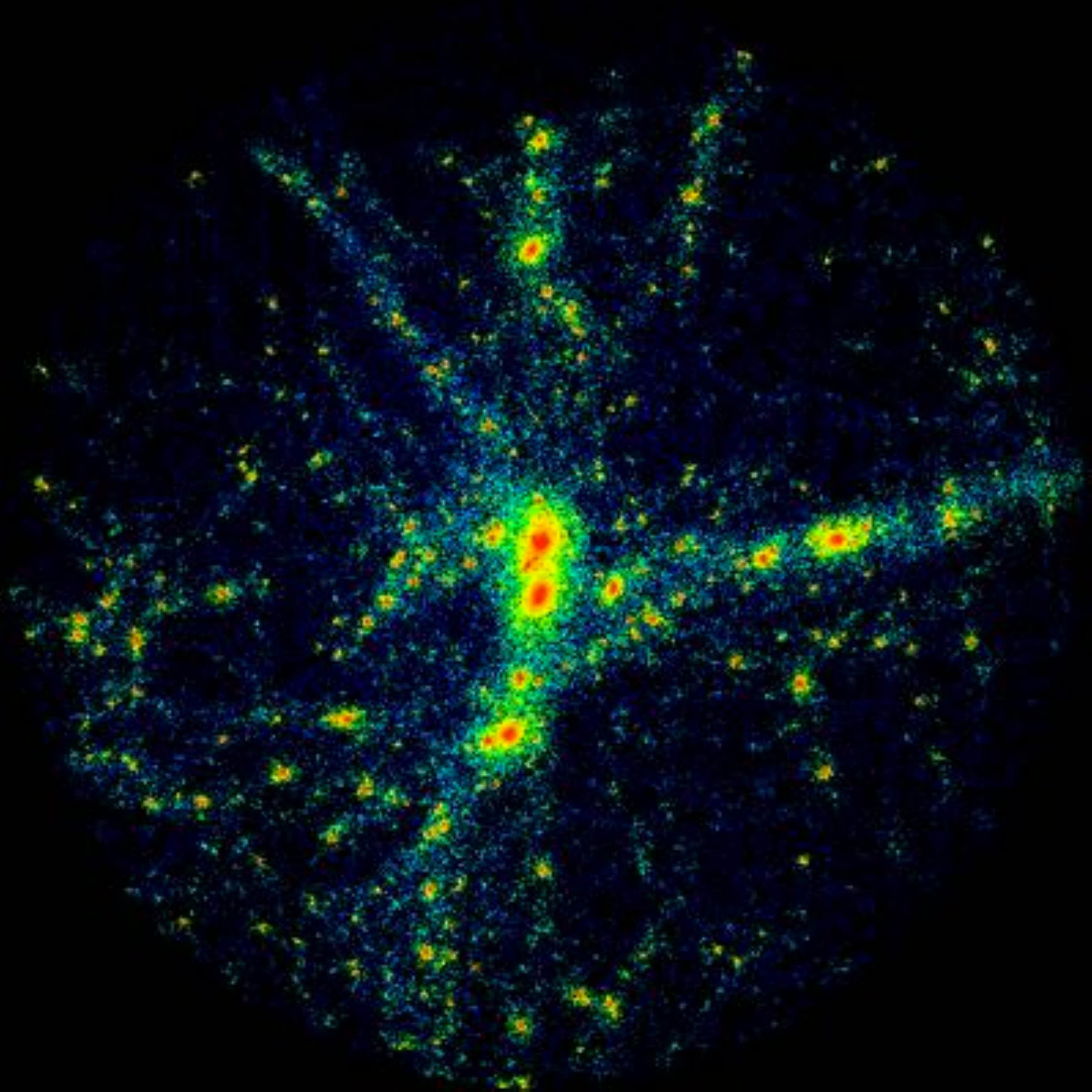}
\hspace{0.24cm}\includegraphics[width=.312\textwidth,trim={-1.cm -0cm -1.cm -0.4cm}]{Figs/haloFHGAzero.pdf}
\caption{\label{fig:model}Baryonic corrections of a halo (of mass $M=10^{14}$ M$_{\odot}$/h) with $f_{\rm bgas}=0.151$, $f_{\rm egas}=0.005$ (left), $f_{\rm bgas}=0.078$, $f_{\rm egas}=0.078$ (middle), and $f_{\rm bgas}=0.005$, $f_{\rm egas}=0.151$ (right). The stellar and dark matter component as well as the ejection radius are kept constant at $f_{\rm cgal}=0.005$, $f_{\rm rdm}=0.839$ and $r_{\rm ej}=3.5\,r_{\rm 200}$ ($\eta_{\rm a}=0.5$). From top to bottom: Density profiles, mass profiles, displacement functions, and cluster density map.}}
\end{figure}

\subsection{Case study of a cluster}
We now illustrate the effects of the baryonic correction model with the example of a randomly selected cluster of mass $M_{200}=10^{14}$ M$_{\odot}$/h, viral radius $r_{\rm 200}=0.77$ Mpc/h, and concentration $c=3.2$. The cluster profile of the initial $N$-body output is well described by an NFW profile plus a constant background component\footnote{Some moderate bumps due to  prominent substructures are visible in the outer part of the profile (see top-panels of Fig.~\ref{fig:model}).}.

We study three hypothetical cases for the cluster with different fractions for the bound and the ejected gas components: (a) nearly all the gas is bound, $f_{\rm bgas}=0.151$, $f_{\rm egas}=0.005$; (b) half of the gas is bound and half is ejected, $f_{\rm egas}=f_{\rm egas}=0.078$; (c) nearly all gas is ejected, $f_{\rm bgas}=0.005$, $f_{\rm egas}=0.151$. For simplicity the dark matter and the stellar fraction are kept constant at $f_{\rm rdm}=\Omega_{\rm c}/\Omega_{\rm m}=0.839$ and $f_{\rm cgal}=0.005$.

The effects due to the different choices for the bound and ejected gas fractions are illustrated in Fig.~\ref{fig:model}, where the density profile, the mass profile, the displacement function, and the density map are plotted from top to bottom. The initial ({\it dmo}) and final ({\it bcm}) model profiles are shown as black and red lines, while the corresponding density and mass measurements from the simulations are given as black and red symbols. Faint non-continues red lines show profiles from the individual components. The displacement function is plotted as dashed line for negative and a solid line for positive values.

Fig.~\ref{fig:model} provides similar information than Fig.~\ref{fig:displ}, however with realistic baryon fractions, making it more difficult to see the baryonic effects on the matter profile. The characteristic density increase in the inner and flattening in the outer part of the profile -- due to gas being condensed into stars and pushed away by feedback effects -- are nonetheless visible. The more gas is in the ejected component, the more the outer density profile is flattened, resulting in an increase of the displacement function. The cluster density maps at the bottom of Fig.~\ref{fig:model} give a visual indication of the changes imposed on the $N$-body outputs. Close inspection allows to observe a slight displacement of satellite haloes in the outer parts of the cluster.

The bottom-line of this section is that the displacement function most crucially depends on the ejected gas fraction. For a cluster with only bound gas the required particle displacement is indeed more than an order of magnitude smaller than for a cluster where all the gas is ejected.

\subsection{Convergence}
It has been shown by previous studies \citep{Heitmann2010,Smith2014,Rasera2014,Heitmann2014,Schneider2015} that both high physical resolution and a large box size are required to guarantee convergence of the power spectrum, resulting in simulations with very high particle numbers. In Ref.~\citep{Schneider2015} we find minimum values $M_p=10^{9}$ M$_{\odot}$/h and $L=500$ Mpc/h for particle number and box size to avoid nonlinear (non Gaussian) deviations from percent convergence. However, these numbers hold for an absolute convergence criterion and do not apply to relative convergence where the difference between a specific model and its corresponding baseline model are examined. It turns out that relative convergence can be achieved with smaller boxes and particle numbers, arguably because box-size and resolution effects are divided out.

\begin{figure}[tbp]
\center{
\includegraphics[width=0.9\textwidth,trim={1.cm 4.5cm 3cm 2cm}]{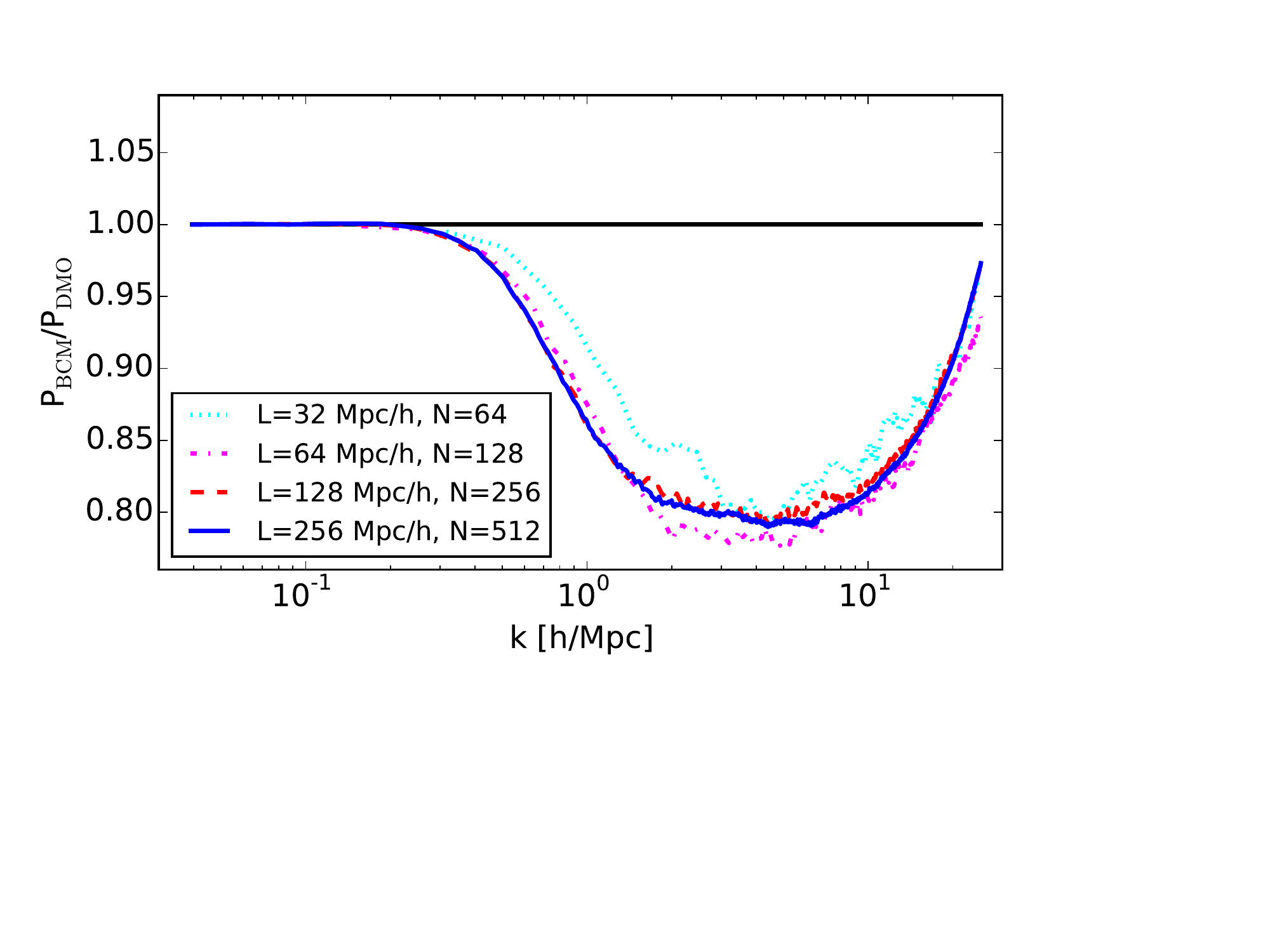}
\caption{\label{fig:PSres}Ratios of power spectra (modified/baseline) from simulation outputs with increasing particle numbers and box sizes and constant physical resolution. Percent convergence is obtained at $N=256$ particles per dimension. The results are based on fiducial model parameters $M_c=1.2\times 10^{14}$ M$_{\odot}$/h, $\beta=0.6$, and $\eta_b=0.5$.}}
\end{figure}

We illustrate the relative convergence properties of the {\it baryonic correction model} in Fig.~\ref{fig:PSres}, where ratios of power spectra from simulations with increasing box-size and particle numbers are shown at fixed physical resolution. While the very low resolution outputs with $N=64$ (cyan) and $N=128$ (magenta) deviate at five and two percent, the $N=256$ (red) result is converged at sub-percent level over all scales with respect to the $N=512$ result (blue). This rapid convergence of the relative power spectrum is very encouraging. It suggests that once a baseline model is simulated to high accuracy with large particle numbers, alternative models can then be simulated at much lower resolution in order to obtain relative deviations to the baseline solution. This is not only true for baryonic effects but should hold for any deviation from the standard model, such as nonlinear effects of massive neutrinos or modifications of gravity.

\section{Baryonic effects on the power spectrum}\label{sec:ps}
In this section we present a detailed study of the effects of baryons on the matter power spectrum. We discuss the influence of individual baryonic components showing the importance of both the fraction and the extend of the ejected gas component. Furthermore, we vary the free parameters of the {\it baryonic correction model} around their fiducial values $M_c=1.2\times 10^{14}$ M$_{\odot}$/h, $\beta=0.6$, and $\eta_b=0.5$. A comparison with X-ray and Sunyaev-Zel'dovich (SZ) observations finally allows us to set constraints on the baryonic power suppression and to come up with a simple, physically motivated fitting function.


\subsection{Single component analysis}
There are different baryonic effects shaping the matter power spectrum at different scales. While the central galaxy and the subsequent adiabatic contraction of the dark matter increases power at small scales, feedback effects such as the AGN activity tend to push the remaining gas away reducing power at larger scales. With the {\it baryonic correction model} it is straight forward to study these effects individually, which allows to gain physical understanding of how baryons shape the matter power spectrum. We do this by treating all but one component as passive non-modified matter components with no influence on the final result.

\begin{figure}[tbp]
\center{
\includegraphics[width=0.9\textwidth,trim={1.cm 4.5cm 3cm 2cm}]{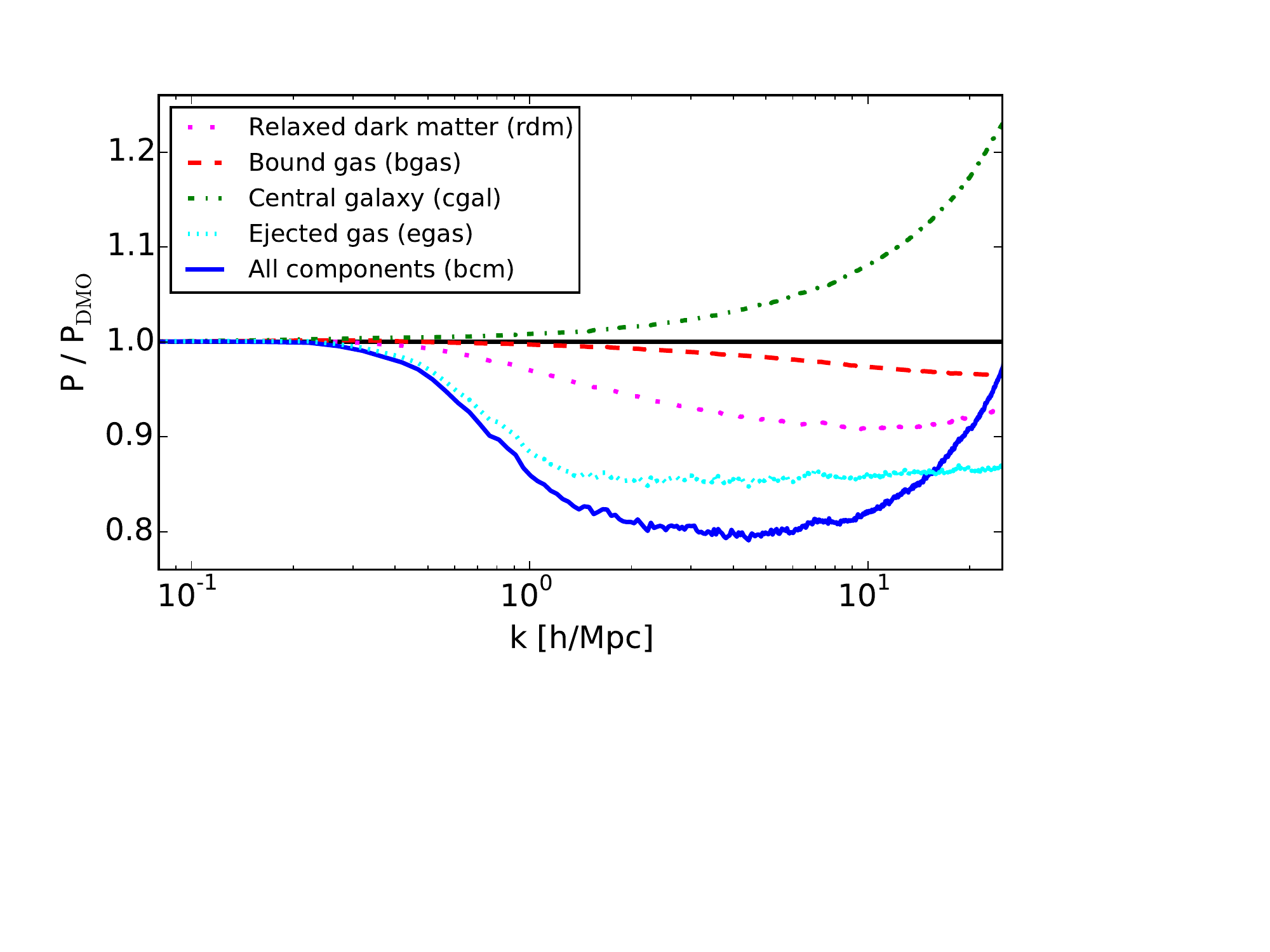}
\caption{\label{fig:PScomp}Power spectra of single components with respect to the DM-only case (black reference line). The total modification (bcm) is illustrated by the solid blue line, while single components are given by the dotted magenta (rdm), dashed red (bgas), dash-dotted green (cgal), and dotted cyan (egas) lines. The results are based on fiducial model parameters $M_c=1.2\times 10^{14}$ M$_{\odot}$/h, $\beta=0.6$, and $\eta_b=0.5$.}}
\end{figure}

The outcome of the single component analysis is illustrated in Fig.~\ref{fig:PScomp}, where the solid blue line shows the total ({\it bcm}) power spectrum and where the individual components correspond to the non-continues coloured lines. In general and as expected, the gas components lead to a power depletion at rather large scales, while the central galaxy increases power at the smallest scales.

The dominant effect on the power spectrum comes from the ejected gas component in form of a significant suppression at small wave numbers. The suppression is further amplified by the adiabatic relaxation of the dark matter which is mainly governed by the missing gas in the halo outskirts. The hot-gas component, on the other hand, only has a very mild effect on the power spectrum yielding an additional suppression of less than three precent. The central galaxy strongly affects the smallest physical scales inducing a significant boost of power at $k>10$ h/Mpc. 

Future galaxy and weak-lensing surveys will probe the matter power spectrum up to $k\sim 1-10$ h/Mpc. It is therefore of prior importance to fully understand the physics of the ejected gas component since it has the strongest effect on the power spectrum at the smallest wave numbers. In the next sections we will show that the ejected gas fraction and the typical ejection radius are key quantities capturing at what scales and how strongly the power spectrum is suppressed.

\subsection{Hot versus ejected: the influence of the gas}\label{sec:PShotvsej}
In the last section we have established that the most stringent modification of the power spectrum comes from the ejected gas. Unfortunately, this is also the baryonic component with the weakest observational constraints, since the ejected gas has a shallow and extended density profile which is very difficult to observe directly. In principle observations of X-ray radiation and the SZ effect can be used to measure gas profiles in clusters and groups. While the X-ray observations predominantly come form the centres of clusters and are ideal to measure the gas fraction inside $M_{500}$, the SZ signal probes the cluster outskirts and therefore the ejected gas profile.

In Sec.~\ref{sec:frac} we have briefly discussed how X-ray observations can be used to determine the fractions of the bound and ejected gas components (the latter indirectly via a subtraction from the cosmic baryon fraction). We now investigate the influence of these components on the matter power spectrum. We do this by following the parametrisation of Eqs.~\eqref{fhga} and \eqref{fega} with the model parameters $M_c$ and $\beta$ describing the mass scale and slope below which (above which) the hot gas fraction (ejected gas fraction) is suppressed. Although the two parameters can be directly constrained with X-ray measurements of the gas fraction at $M_{500}$, there is significant uncertainty due to statistical and potentially systematic errors. We therefore allow the parameters to freely vary within a reasonable parameter range (enclosing the observations) and study their effects on the power spectrum.

\begin{figure}[tbp]
\center{
\includegraphics[width=0.47\textwidth,trim={1cm 0cm 3cm 1.0cm}]{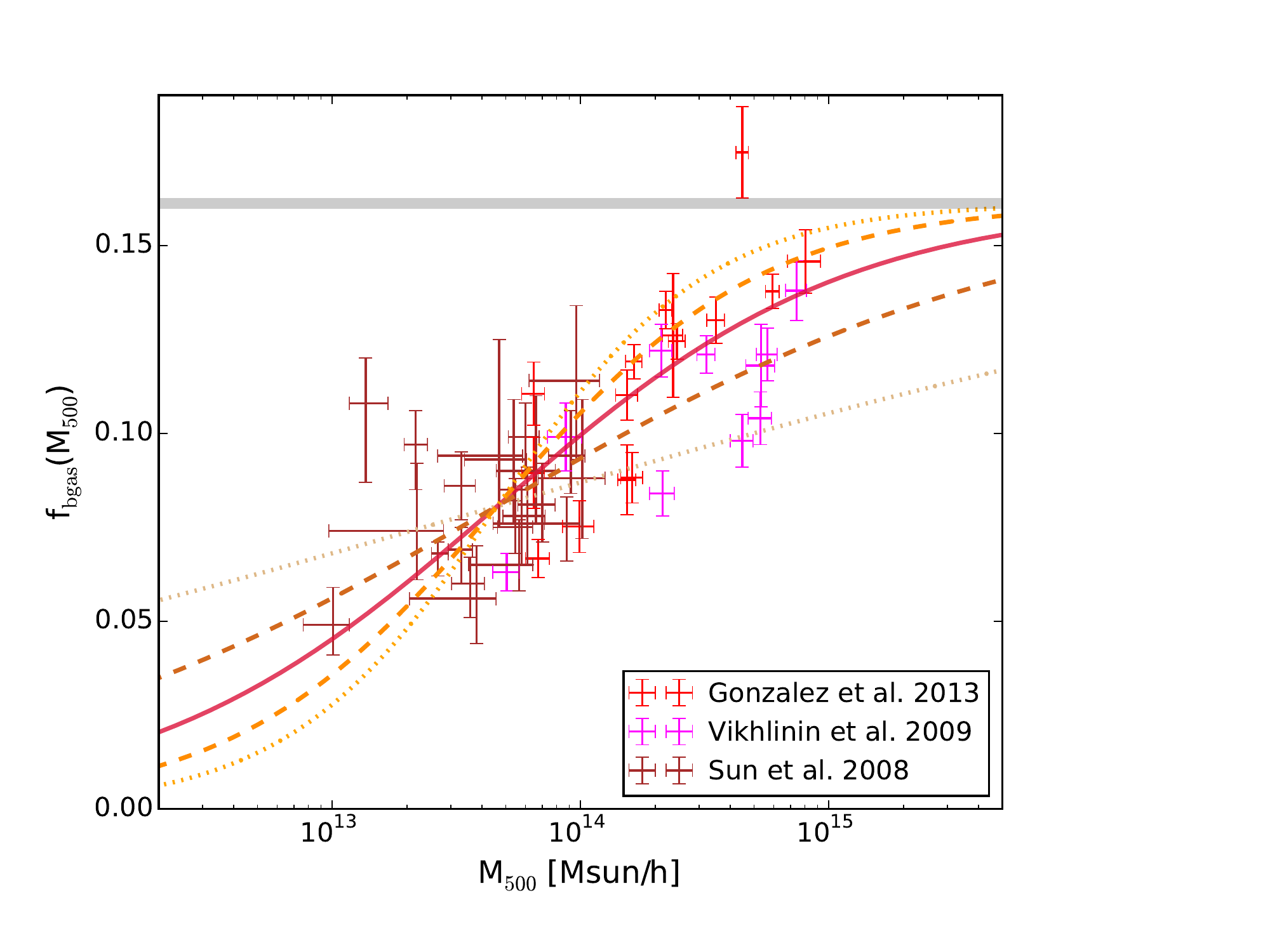}
\includegraphics[width=0.47\textwidth,trim={1cm 0cm 3cm 1.0cm}]{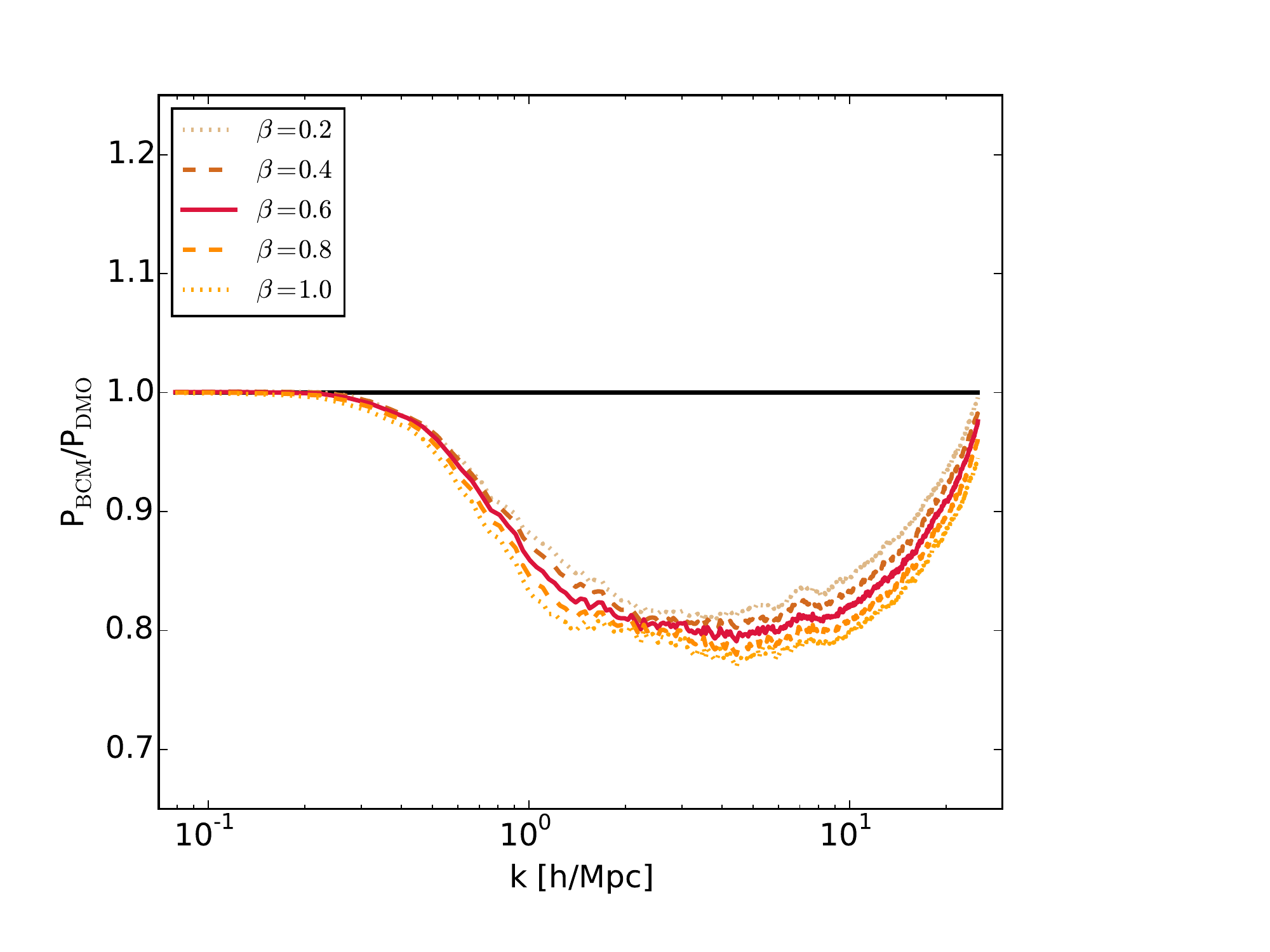}\\
\includegraphics[width=0.47\textwidth,trim={1cm 1cm 3cm 1.0cm}]{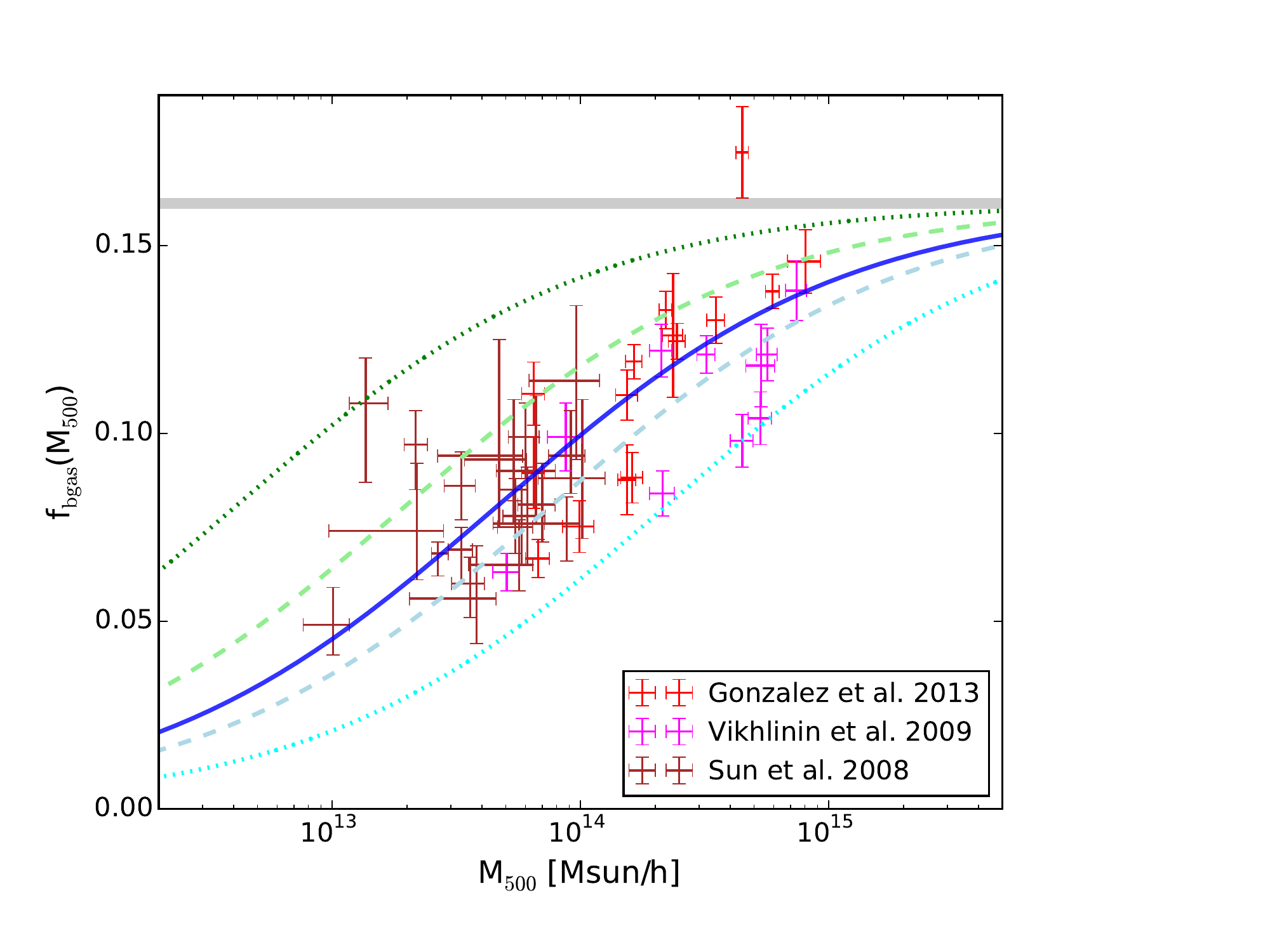}
\includegraphics[width=0.47\textwidth,trim={1cm 1cm 3cm 1.0cm}]{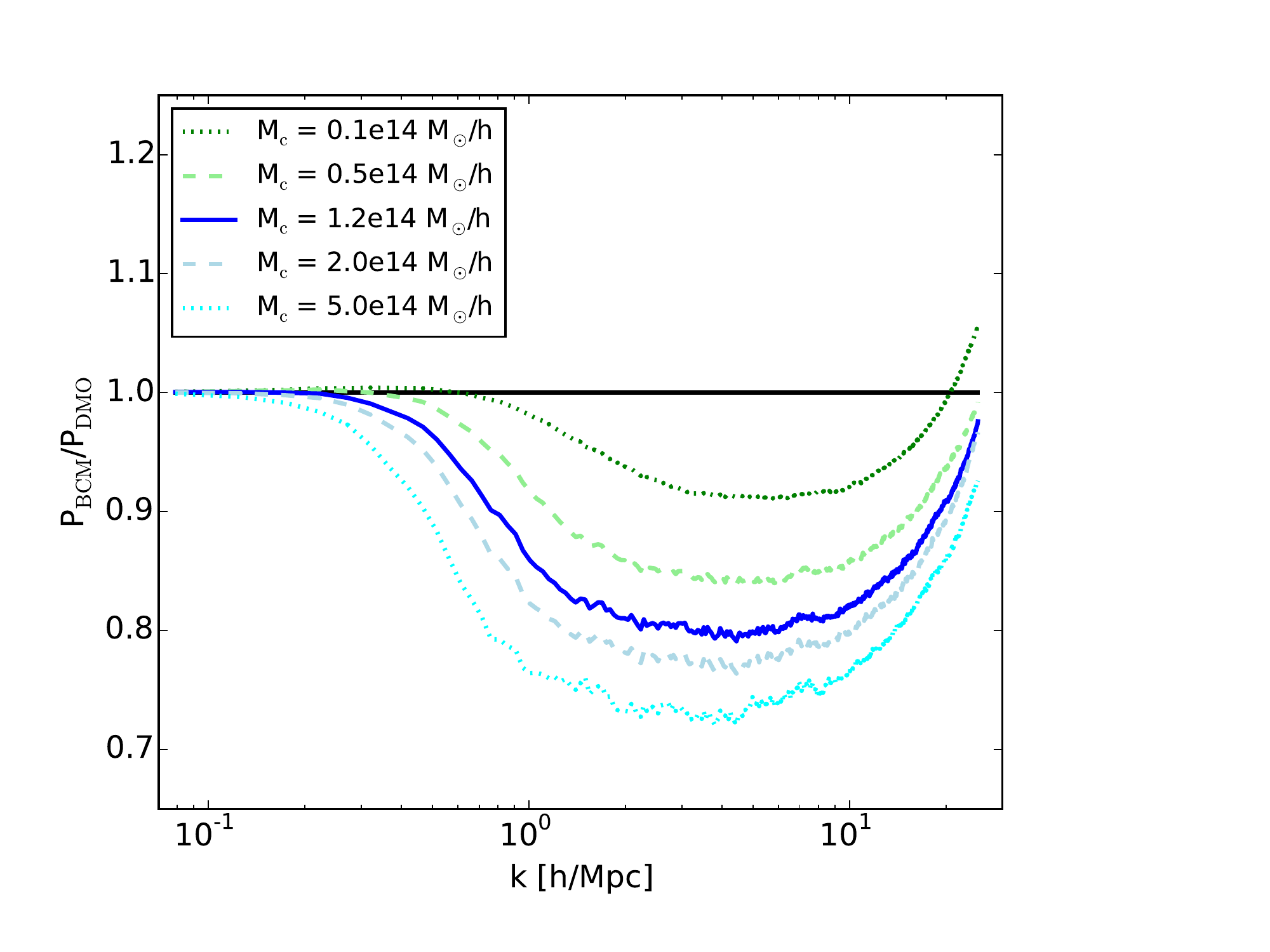}
\caption{\label{fig:PS_Mcbeta}Left: different choices of the fitting parameters from Eq.~\eqref{fhga}: $\beta$ (top) and $M_c$ (bottom). Right: Ratio of the Power spectrum with corresponding parametrisation.}}
\end{figure}

In the top panels of Fig.~\ref{fig:PS_Mcbeta} we illustrate how the variation of $\beta$ affects the matter power spectrum. The different values investigated are $\beta=$ 0.2, 0.4, 0.6, 0.8, 1.0 (at fixed parameter $M_c=1.2\times 10^{14}$ M$_{\odot}$/h) with the best fitting value $\beta=0.6$ highlighted as a solid red line. Surprisingly, the power spectrum shows very little variation for changes in the $\beta$-parameter, staying within three percent over the entire range of $k$-modes\footnote{We have checked that $\beta$ still has a minor effect on the power spectrum if $M_c$ is changed to 0.5 or $2\times10^{14}$ M$_{\odot}$/h. However, for an extreme value of $M_c$ outside of the mass range $\sim10^{13}-5\times10^{14}$ M$_{\odot}$/h (most relevant for the power spectrum) one expects the $\beta$-parameter to have a much stronger effect.}.

The effect of the second parameter $M_c$ is illustrated in the bottom panels of Fig.~\ref{fig:PS_Mcbeta}. Here we examine the values $M_c=0.1$, 0.5, 1.2, 2.0, $5.0\times 10^{14}$ M$_{\odot}$/h (with fixed value $\beta=0.6$). The fiducial value $M_c=1.2\times 10^{14}$ M$_{\odot}$/h, providing the best fit to the observations, is highlighted as solid blue line. Contrary to $\beta$, the choice of $M_c$ strongly affects the power spectrum. Within the probed range the total power suppression grows from roughly 10 to 25 percent for increasing values of $M_c$.

The results presented in Fig.~\ref{fig:PS_Mcbeta} are encouraging because they show that the suppression of the power spectrum mainly depends on the critical mass and not so much on the slope of the ejected/hot gas-mass relation (i.e. Eq.~\ref{fhga}). This makes it easier to find a physically motivated fitting function connecting key values of the gas distribution to the power spectrum suppression (see Sec.~\ref{sec:fit}).

\subsection{Ejection radius: effects at the largest scales}
The profile of the ejected gas component can only be determined with observations from the outskirts of haloes. Recently it has become possible to constrain the gas fraction beyond the viral radius with SZ measurements from {\it Planck} \citep{Ade2013}. These results allow to tentatively distinguish between different values for the ejected gas radius ($r_{\rm ej}$) at cluster scales. For smaller haloes, $r_{\rm ej}$ can be determined with either model (A) or (B) defined in Sec.~\ref{sec:rcr}. While model (A) simply assumes a constant fraction $r_{\rm ej}/r_{200}$ over all mass ranges, model (B) accounts for the fact that gas is more efficiently ejected out of smaller haloes. We mainly focus on model (B) which is physically motivated and constrain the corresponding model parameter $\eta_b$. However, further observations of the gas fraction around galaxy groups and small clusters are crucial in order to confirm the validity of model (B) and its predictions with respect to the power spectrum. 

\begin{figure}[tbp]
\center{
\includegraphics[width=0.47\textwidth,trim={1cm 0cm 3.cm 1cm}]{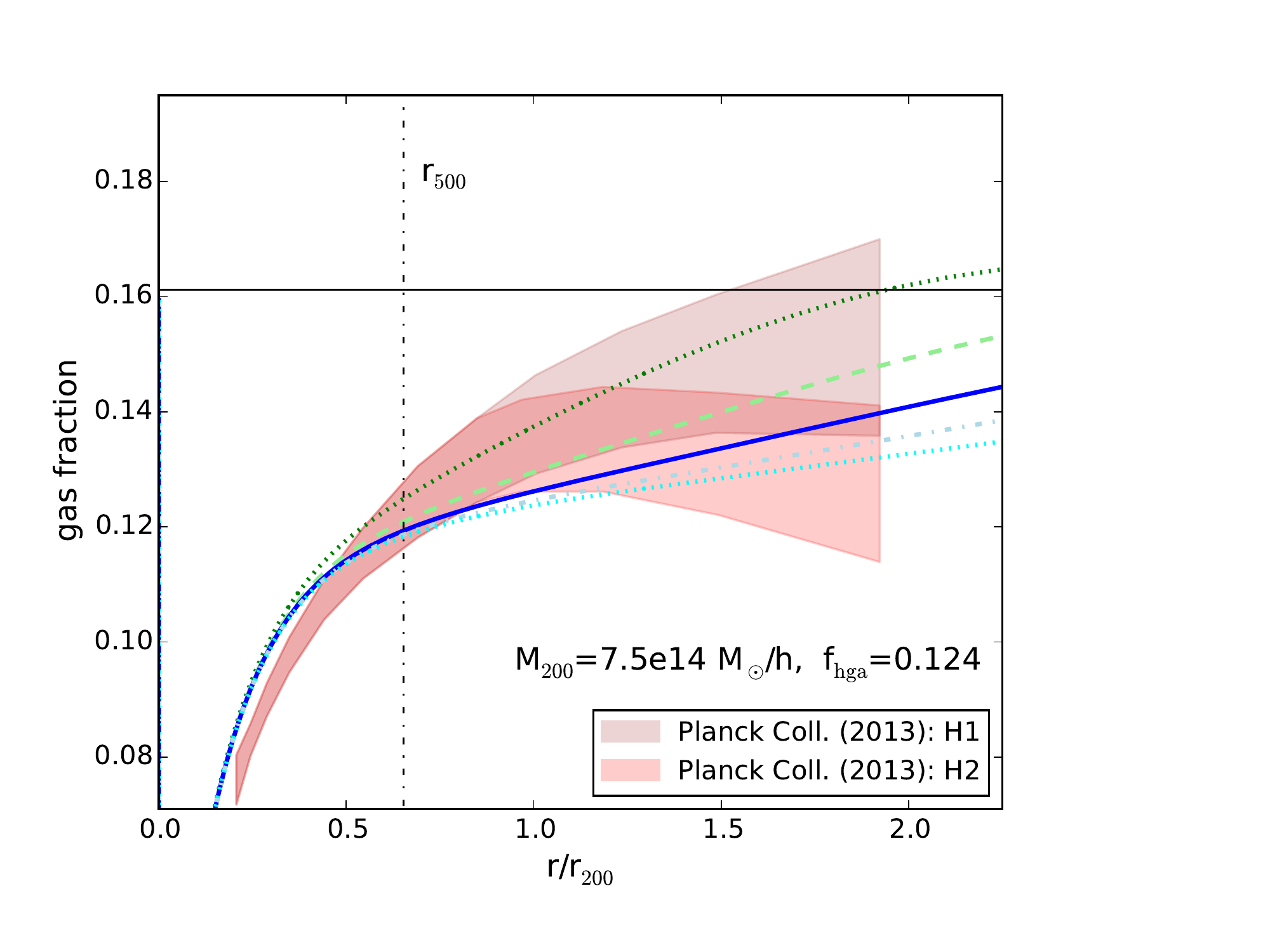}
\includegraphics[width=0.47\textwidth,trim={1cm 0cm 3.cm 1cm}]{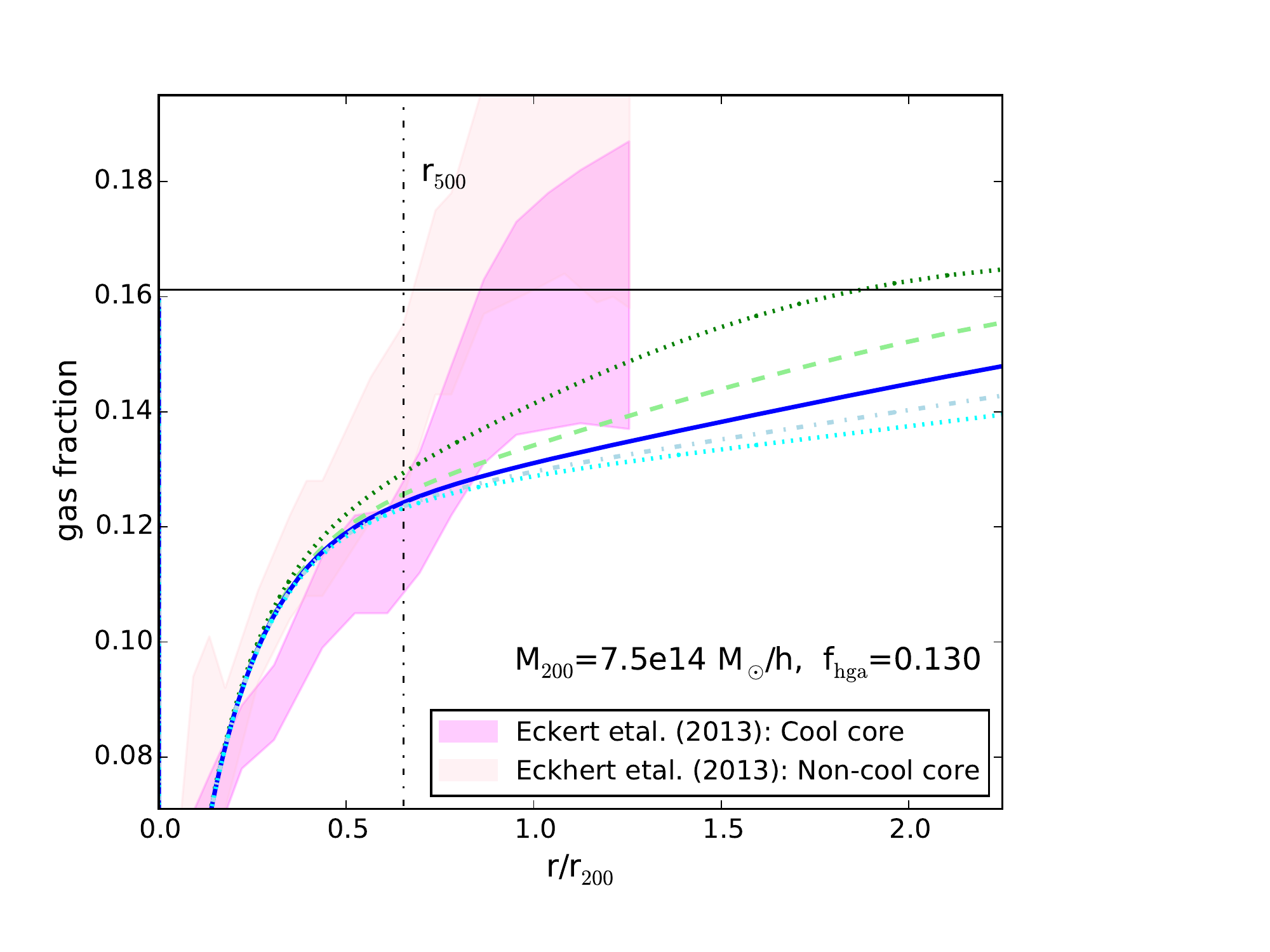}\\
\includegraphics[width=0.47\textwidth,trim={1cm 1cm 3.cm 1cm}]{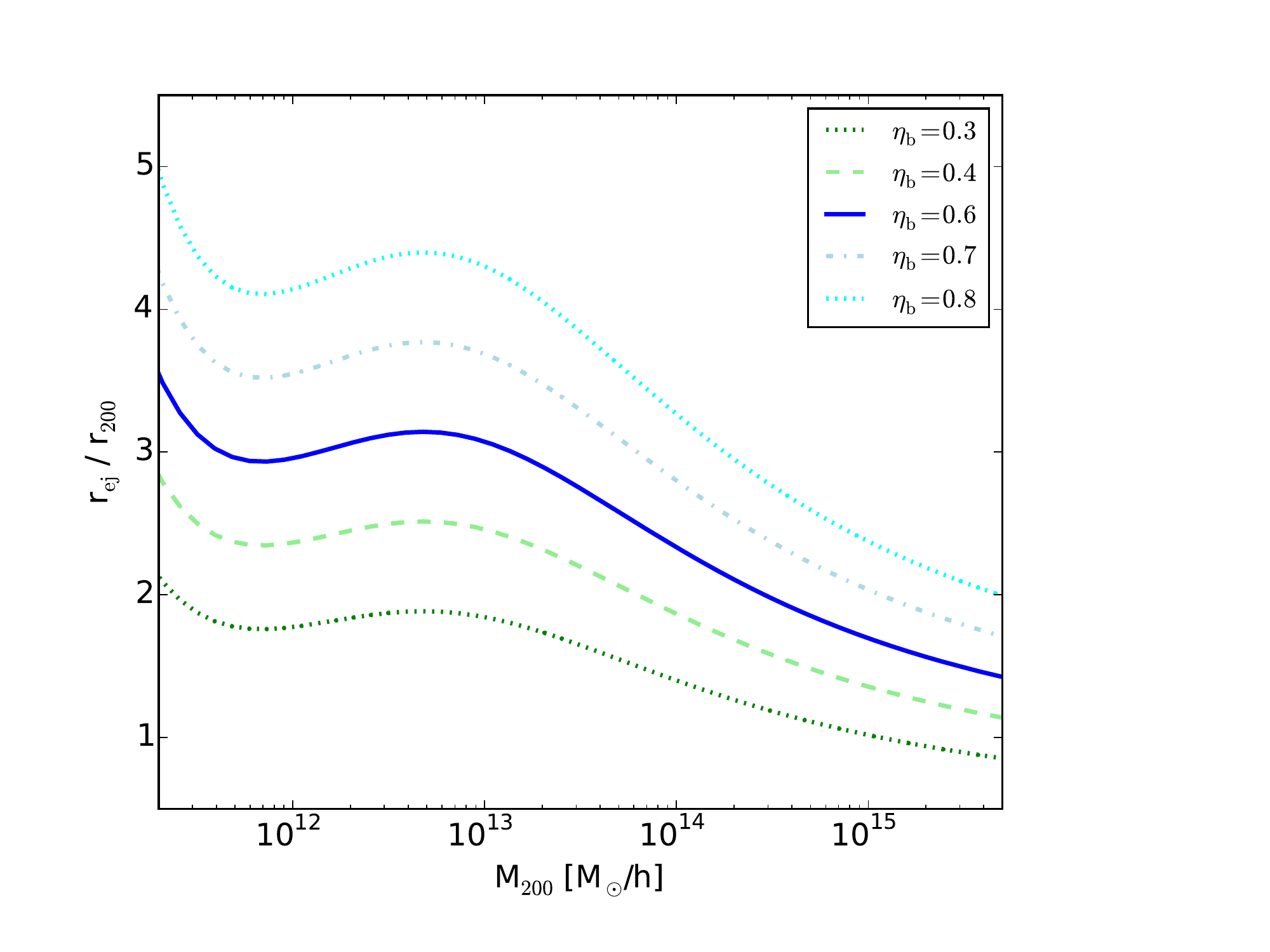}
\includegraphics[width=0.47\textwidth,trim={1cm 1cm 3.cm 1cm}]{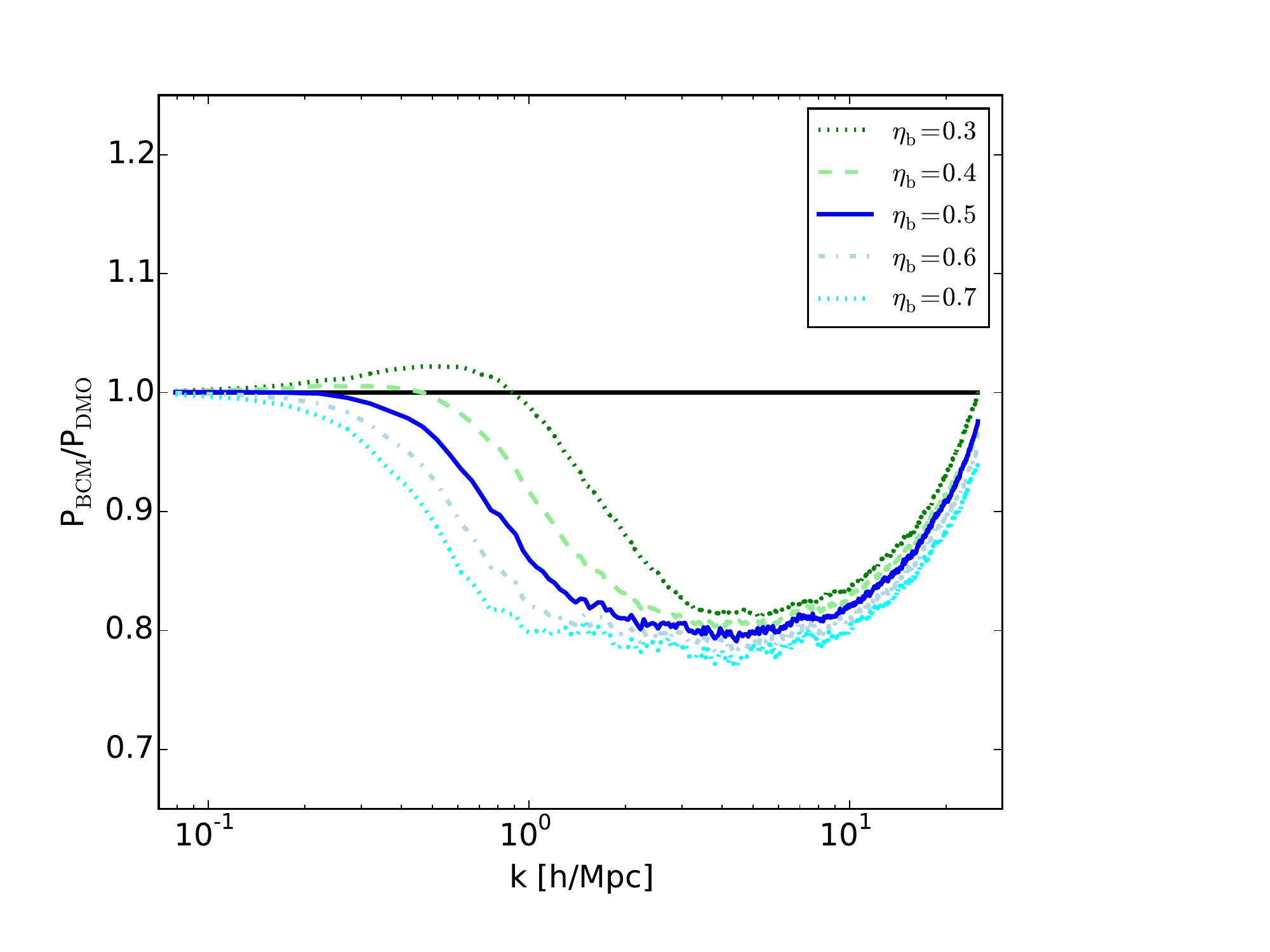}
\caption{\label{fig:PS_rej}{\it Top:} Radial profile of the cluster gas fraction from the {\it Planck} collaboration \citep{Ade2013} (left, with different approximation models H1 and H2) and from Ref. \citep{Eckert2013} (right, separated by clusters with cool and non-cool cores) compared to the {\it baryonic correction model} with different values for $\eta_b$. {\it Bottom:} Mass dependence of the ejection radius from model (B) of Sec.~\ref{sec:rcr} (left) and resulting power spectra (right)}}
\end{figure}

In the upper panels of Fig.~\ref{fig:PS_rej}, we plot the gas fraction from the {\it baryonic correction model} together with observational results based on {\it Planck} SZ data. The left panel shows measurements from the {\it Planck} collaboration \citep{Ade2013}, who combined SZ pressure profiles with extrapolated X-ray data from {\it XMM-Newton}. Different extrapolation schemes yield different results (labelled H1 and H2 in the figure, see Ref.~\citep{Ade2013} for more information) which were claimed to roughly bound the true profile \citep{Battaglia2013}. The results are in general agreement with our model, however with too large error bars to show clear preferences for a certain value of $\eta_b$. While model H1 prefers smaller ejection radii (i.e smaller $\eta_b$), model H2 seems very inconclusive.

The right panel of Fig.~\ref{fig:PS_rej} shows results from Ref.~\citep{Eckert2013} based on SZ measurements from {\it Planck} and {\it ROSITA} data, using a different mass estimate based on the assumption of hydrostatic equilibrium. This leads to a generally higher gas fraction around the viral radius, especially for non-cool core clusters, pointing towards rather small ejection radii with $\eta_b<0.5$. Very recent results solely based on X-ray observations from {\it Chandra} confirm these rather high gas fractions~\citep{Morandi2015}. However, Ref.~\citep{Battaglia2013} argue that assuming hydrostatic equilibrium plus not resolving small gas clumps may artificially boost the observed gas fraction.

In Appendix \ref{app:sim} we compare the gas fraction profile of our model to hydrodynamical simulations from Ref.~\citep{Battaglia2013} and find good agreement at all investigated mass scales. The best match is obtained for $\eta_b\sim 0.6$ pointing towards efficient feedback in the simulation in slight tension with the observational results presented above. The good agreement between the simulations and the {\it baryonic correction model} further confirms the validity of the assumed gas profiles.

The bottom panels of Fig.~\ref{fig:PS_rej} illustrate the ejection radius for a given value of $\eta_b$ (left) and the corresponding ratios of the power spectrum (right). The mass dependence of the ejection radius results from the fact that small haloes have shallower potentials allowing the ejected gas to expand further as described by model (B) in Sec.~\ref{sec:rcr}. The power spectra shown on the right-hand-side of Fig.~\ref{fig:PS_rej} are sensitive to the choice of the ejected gas radius. Increasing the value of $\eta_b$ leads to a shift of the power suppression towards smaller $k$-modes (larger physical scales) while there is only a minor effect on the maximum power suppression. Hence, there is a simple trend between the ejection radius of gas around haloes and the largest scale where baryonic effects become important\footnote{The simpler model (A), which is based on constant fractions $r_{\rm ej}/r_{200}$, yields nearly indistinguishable power spectra for the values $\eta_a=0.75\,\eta_b$ (see Sec.~\ref{sec:rcr}). This corresponds to $r_{\rm ej}/r_{200}=1.6,2.2,2.7,3.2,3.8$ for the values of $\eta_b$ investigate in Fig.~\ref{fig:PS_rej}.}.

Based on the established connection between power suppression and ejection radius, it is possible to estimate the largest scale where baryons affect the power spectrum. The SZ based measurements from Ref.~\citep{Eckert2013} prefer values of $\eta_b\lesssim0.4$ resulting in baryonic effects being negligible for modes $k<0.5$ h/Mpc. Alternative modelling by Ref.~\citep{Ade2013} confirms this result for one (but not the other) approximation scheme. Further confirmation comes from direct X-ray measurements by Ref. \citep{Morandi2015} which also point towards small small ejection radii for clusters.

Although it is encouraging that current observations are able to constrain the model parameters of the ejected gas profile, the limits on the power suppression have to be taken with a pinch of salt. Indeed, there are two main caveats potentially affecting the result: First, there could be a bias because of unresolved gas clumping and the assumption of hydrostatic equilibrium, artificially boosting the observed values of the gas fraction \citep{Battaglia2013}. Second, all current observations come from large galaxy clusters while the power spectrum is mainly influenced by group-size haloes. We use a physically motivated model to estimate the ejection radius-mass relation based on the cluster results, which ultimately requires observational confirmation. For the future, it is therefore crucial to find ways to measure gas profiles of Milky-Way to group sized objects in order to better predict the mass clustering of the universe.

\subsection{The role of scatter}\label{sec:scatter}
One of the advantages of the {\it baryonic correction model} is its adaptability to different approximations or new model assumptions. For example, it is straight forward to introduce a scatter term for the bound and ejected gas fractions. The observational data of Fig.~\ref{fig:fhga} indeed shows a significant amount of scatter. Some of it should come from uncertainties in the data, but there should also be a physical contribution due to the fact that haloes have individual formation histories affecting the gas components.

In order to test the influence of scatter in the bound and ejected gas components, we add Gaussian scatter to the bound gas fraction ($f_{\rm bgas}$). This means that, for every halo in the simulation output, $f_{\rm bgas}$ is drawn randomly from a Gaussian distribution with standard deviation  $\sigma=0.1,\,0.2$ around the mean given by Eq.~\eqref{fhga}.

\begin{figure}[tbp]
\center{
\includegraphics[width=0.47\textwidth,trim={1cm 1cm 3cm 1cm}]{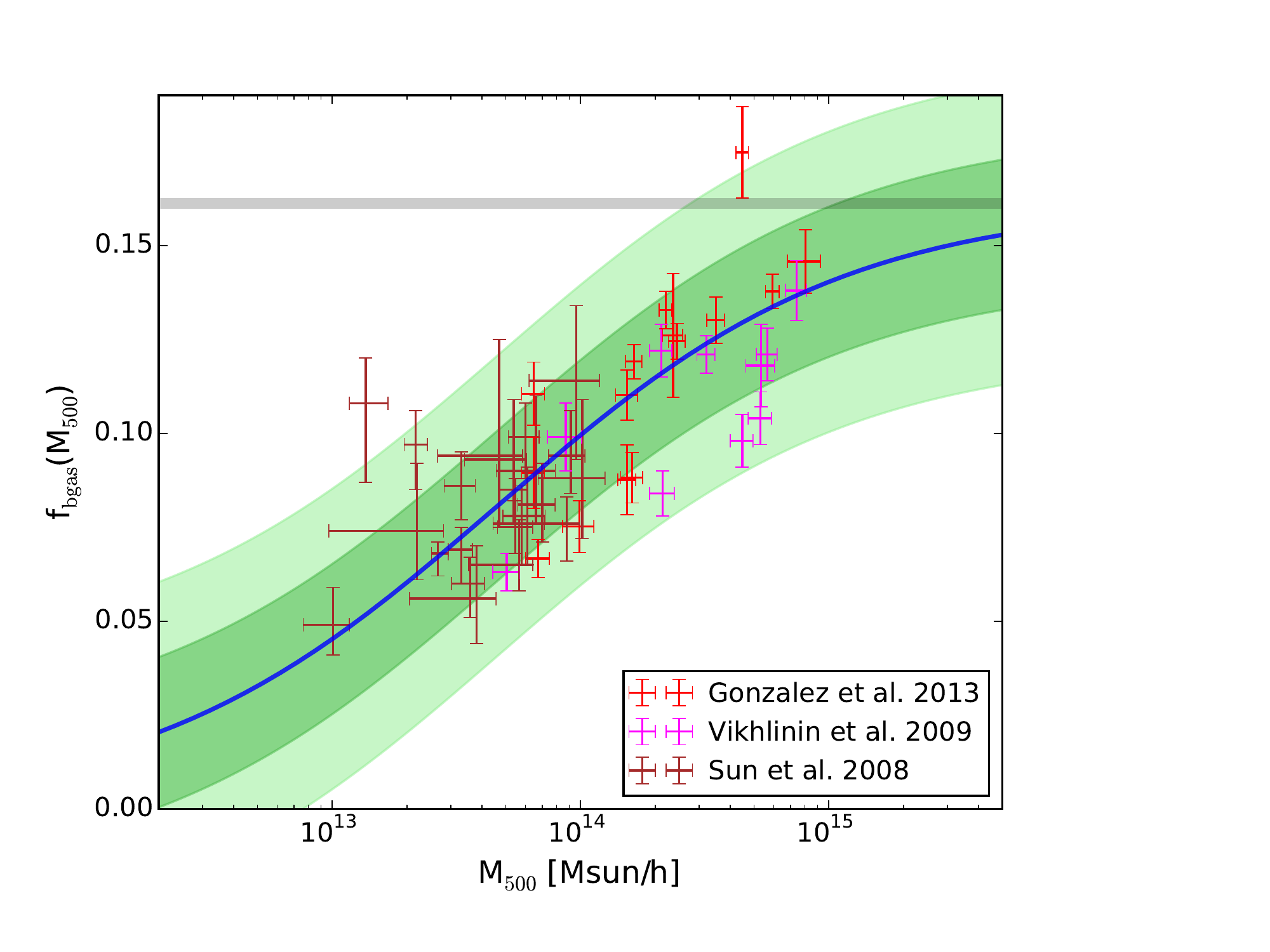}
\includegraphics[width=0.47\textwidth,trim={1cm 1cm 3cm 1cm}]{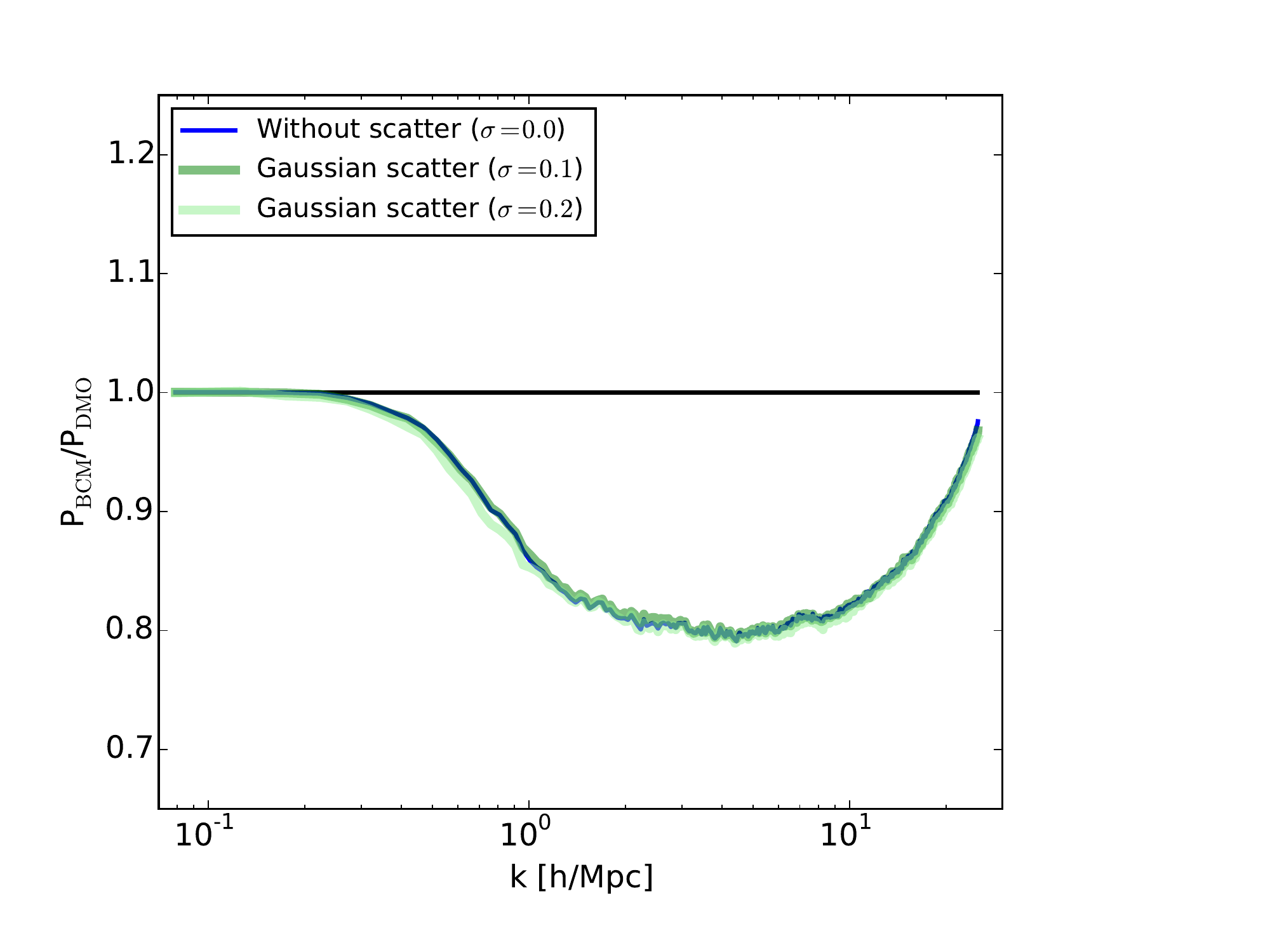}
\caption{\label{fig:PS_scatter}Left: 2-$\sigma$ contours of a Gaussian scatter model with $\sigma=0.1$ (green) and $\sigma=0.2$ (light green) around the bound gas fraction $f_{\rm bgas}(M)$ with the best-fitting parameters ($M_{\rm c}=1.2\times10^{14}$ M$_{\odot}$/h, $\beta=0.6$). Right: corresponding ratios of the power spectrum.}}
\end{figure}

The influence of the scatter on the power spectrum is shown in Fig.~\ref{fig:PS_scatter}. While the left panel illustrates the 2-$\sigma$ scatter-contours around the best fitting function for the bound gas fraction, the right panel shows how the power spectra is affected. Despite the significant scatter of $\sigma=0.1$ (green) and $\sigma=0.2$ (light green), the power spectrum stays nearly identical to the scenario without scatter (blue). This is very encouraging, as it shows that individual gas fractions from the specific halo formation histories do not influence the matter power spectrum significantly, strengthening the results presented here.

\subsection{Redshift dependence}
The effects of baryons on the matter power spectrum are intrinsically redshift dependent. At high redshift, prior to the typical formation time of group-sized haloes, the effects are negligible, but later on, they are expected to grow steadily due to repeated gas ejection via feedback mechanisms.
Quantifying the redshift dependence is especially important for predicting the cosmic shear because the signal depends on the power spectrum between roughly redshift zero and two depending on the distribution of sources (see e.g. \citep{Laureijs2011}).

Implementing the full redshift dependence into the baryonic correction model might appear challenging, as parameters like $M_c$ and $\eta_b$ could vary with redshift. Pinning down the redshift dependence of these parameters would require high-redshift observations of the halo gas fraction, which currently do not exist. However, AGN activity is observed to peak around $z=2.15$ \citep{Hopkins2006}, suggesting the the model parameters $M_c$ and $\eta_b$ might not evolve much below redshift two.

It is important to realise that the baryonic correction model leads to intrinsically redshift dependent power spectra even for the case of constant model parameters. This is due to the fact that at different redshifts the power spectrum measurement is dominated by haloes of different mass scales. For example, small clusters strongly contribute to the signal at redshift zero, but they are too rare to have an influence above redshift one. In this paper we study the redshift dependence (of the range $z=0-2$) due to the growth of haloes while assuming redshift independent model parameters. This rather strong assumption ultimately requires testing with simulations and, ideally, observations, an endeavour we postpone to future work.

\begin{figure}[tbp]
\center{
\includegraphics[width=0.47\textwidth,trim={1cm 0cm 3.cm 1cm}]{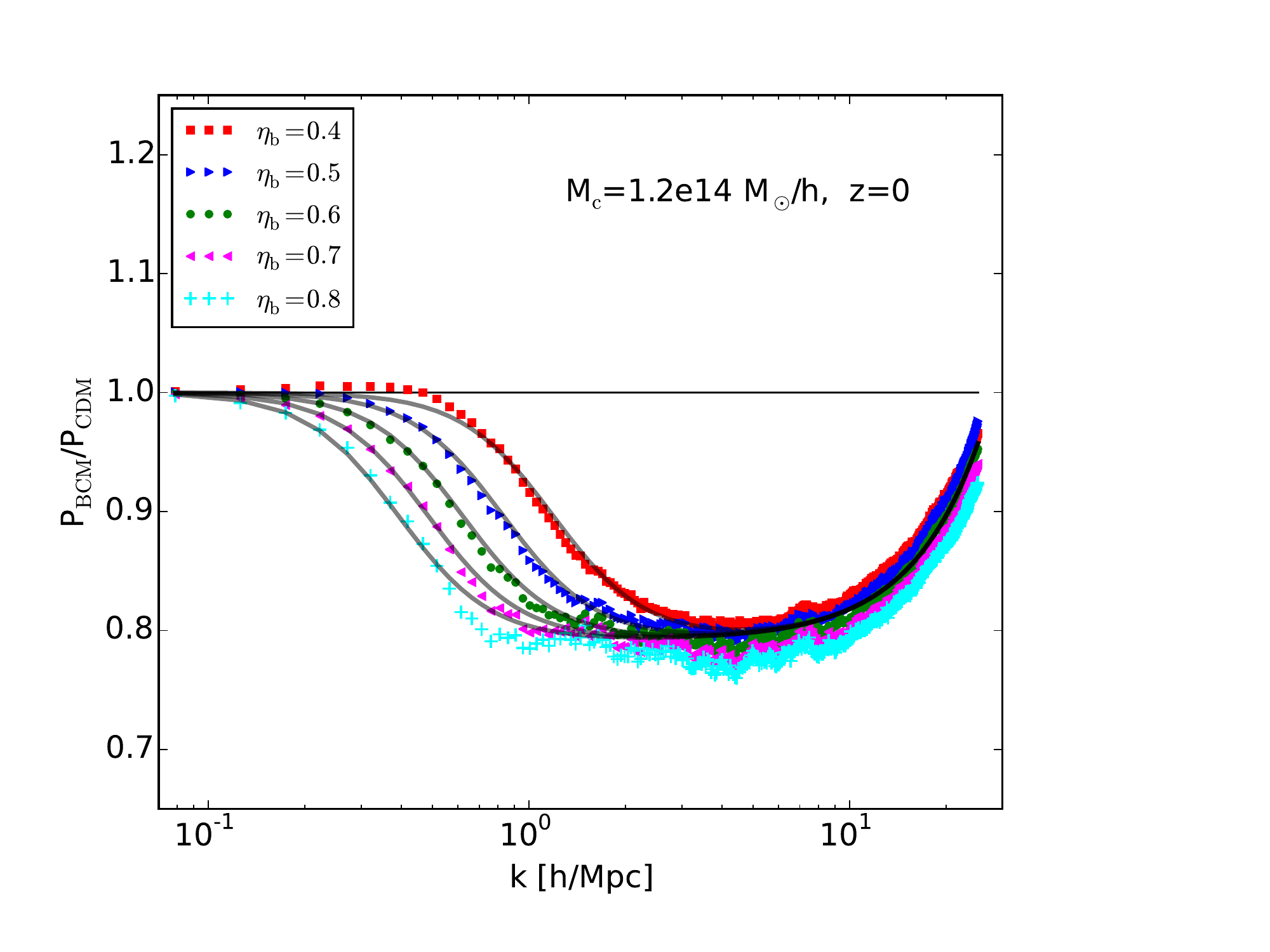}
\includegraphics[width=0.47\textwidth,trim={1cm 0cm 3.cm 1cm}]{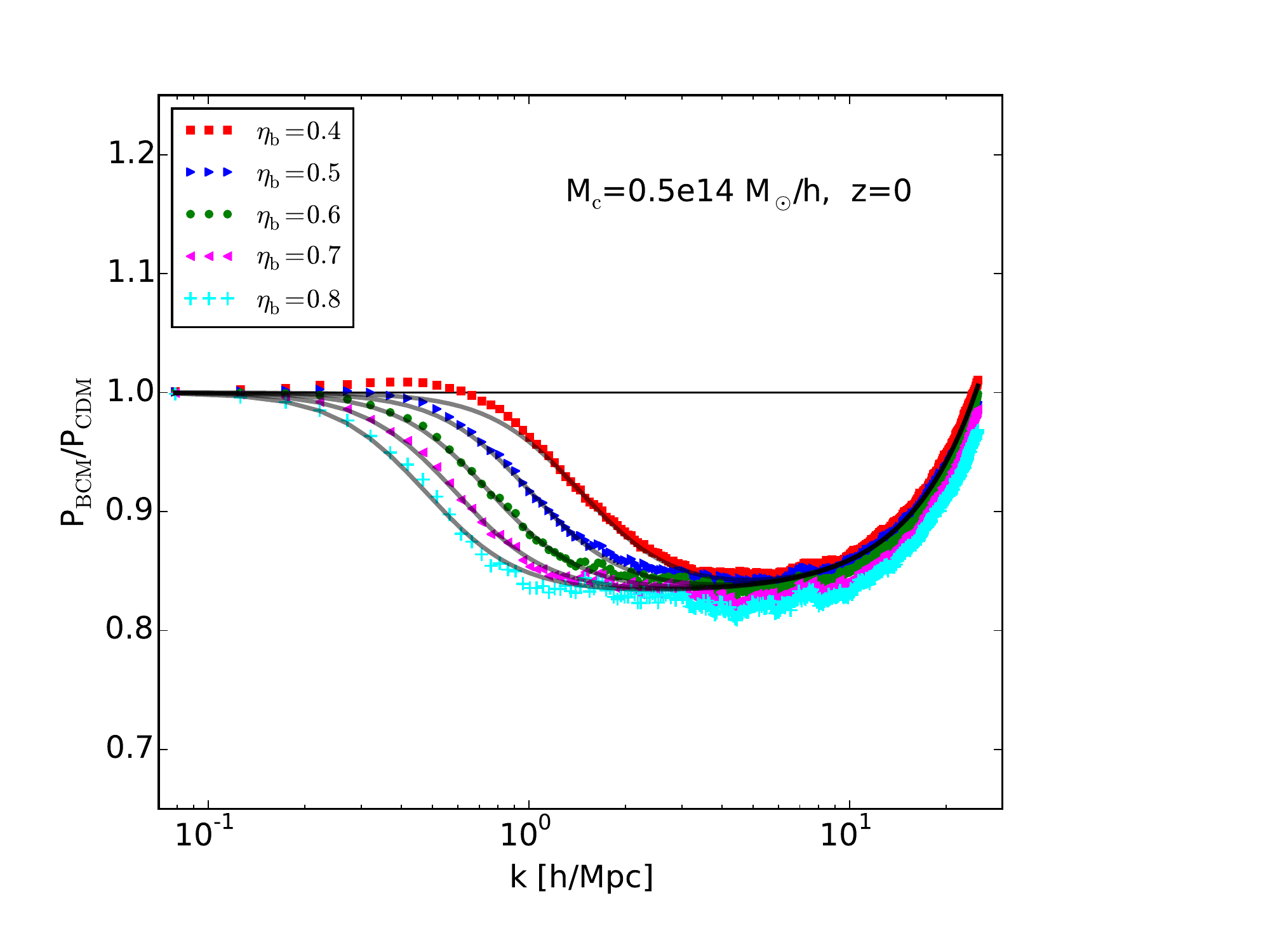}\\
\includegraphics[width=0.47\textwidth,trim={1cm 0cm 3.cm 1cm}]{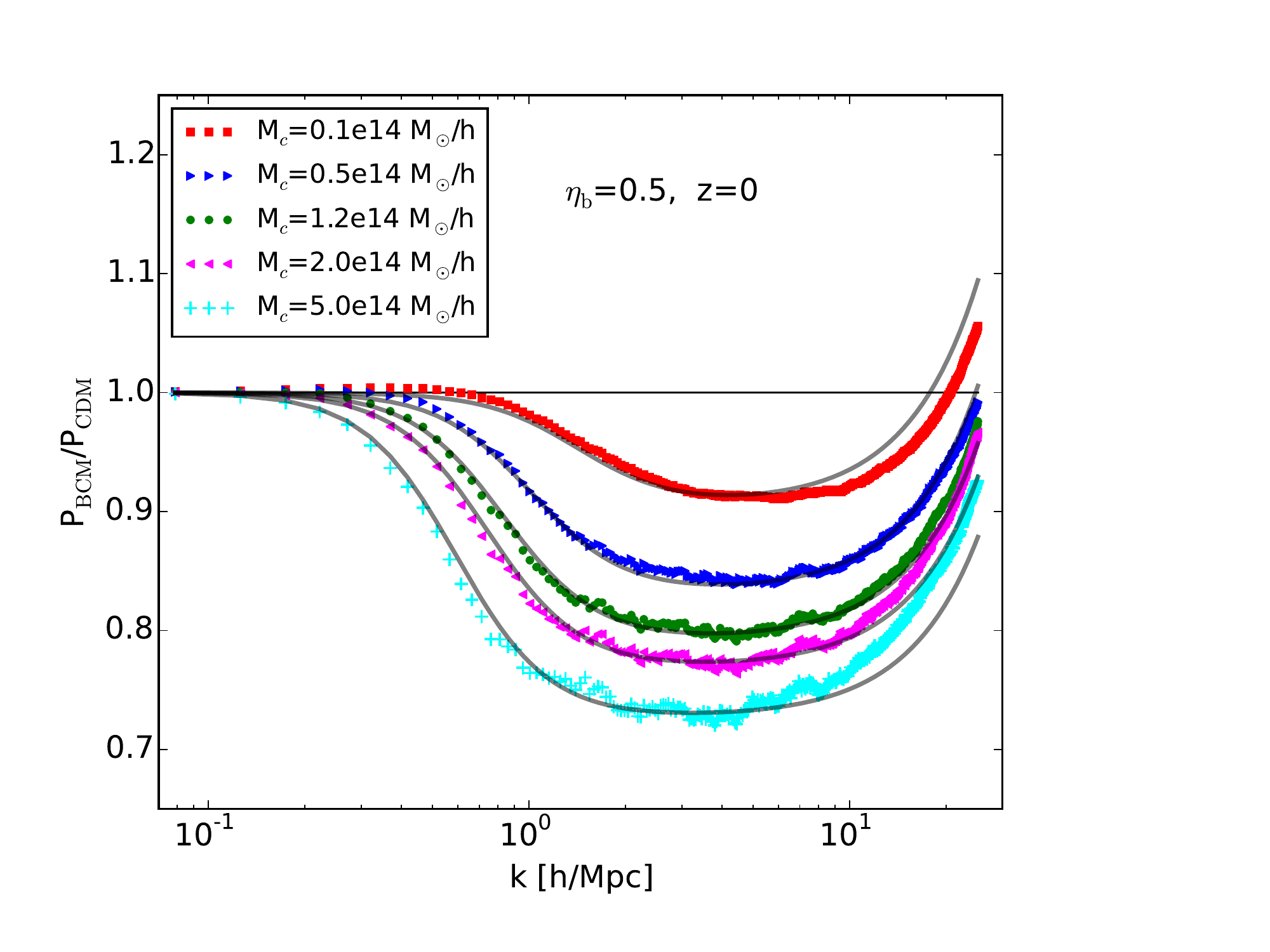}
\includegraphics[width=0.47\textwidth,trim={1cm 0cm 3.cm 1cm}]{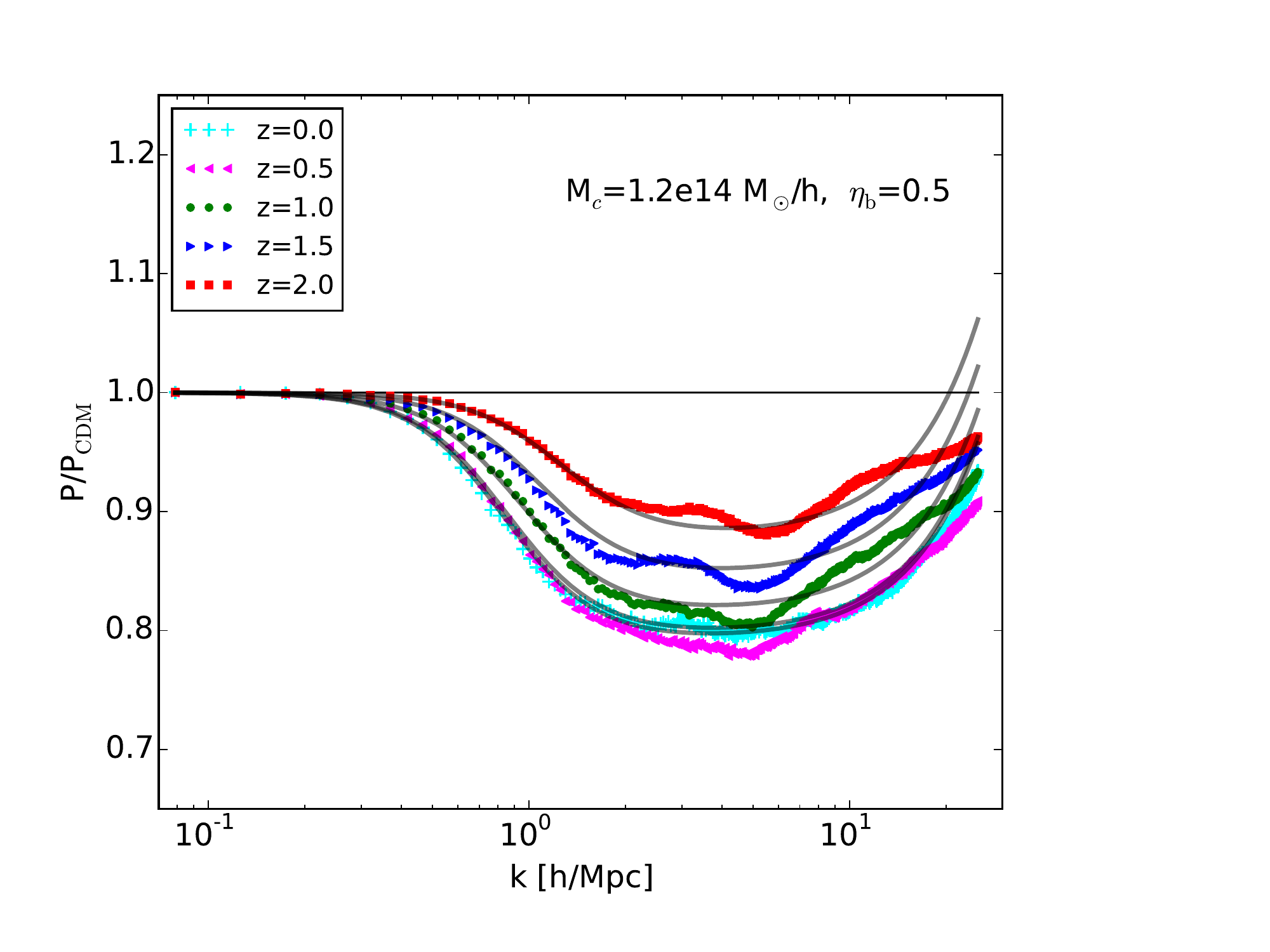}
\caption{\label{fig:PSfit}Selection of measured power spectra with varying parameters $\eta_{\rm b}$ (top), $M_{\rm c}$ (bottom left),  and redshift (bottom right). Coloured symbols are measurements form simulation outputs, grey lines represent the fitting function of Eq.~\eqref{fit}.}}
\end{figure}

The bottom-right panel of Fig.~\ref{fig:PSfit} shows the redshift dependence of the power spectrum, illustrating how the baryonic suppression grows substantially between $z\sim 2-1$, slows down at $z=1$, and stops to evolve around $z=0.5$. This qualitative evolution is in agreement with the redshift dependence observed in hydrodynamical simulations of Ref.~\citep{Schaye2010} (but see Fig.~1 in Ref.~\citep{Semboloni2011}), where the growth of suppression is shown to be significant at $z=2-1$ and small at $z=1-0$. In both our model and the simulations the total baryonic suppression roughly doubles in scale between redshift two and zero, an encouraging agreement supporting our initial assumption of redshift independent model parameters.

\subsection{Fits for the power spectrum with baryons}\label{sec:fit}
In the last sections we have established that the presence of baryons leads to both a power suppression at medium and a power increase at small scales, the former due to gas ejection and the latter due to the central galaxy. The total amount of suppression is driven by the typical halo mass scale ($M_{\rm c}$) below which most of the gas is ejected, while the maximum scale (minimal $k$-range) where the suppression becomes visible depends on the parameter $\eta_b$ (fixing the $r_{\rm ej}$-mass relation). Furthermore, the suppression is growing with decreasing redshift as the signal is dominated by larger and larger haloes. These simple trends make it possible to determine a fitting function based on three parameters, $M_{\rm c}$, $\eta_b$ and $z$, and relating key observables of the gas distribution in haloes to the shape of the power spectrum. The fitting function has the following functional form:
\begin{equation}\label{fit}
F(k,z) = \frac{P_{\rm BCM}}{P_{\rm DMO}}=G(k|M_c,\eta_b,z)S(k|k_s),
\end{equation}
where $G$ describes the suppression due to the gas and $S$ the small-scale increase due to the central galaxy stars. The gas suppression is best captured by the function
\begin{equation}
G(k|M_c,\eta_b,z)=\frac{B(z)}{1+(k/k_{g})^3}+\left[1-B(z)\right]
\end{equation}
where
\begin{equation}
B(z)=B_0\left[1+\left(\frac{z}{z_c}\right)^{2.5}\right]^{-1},\hspace{1cm}k_g(z)=\frac{0.7\,\left[1-B(z)\right]^4\,\eta_b^{-1.6}}{\rm h/Mpc},
\end{equation}
with $z_c=2.3$ and
\begin{equation}
B_0=0.105\log_{10}\left[\frac{M_{c}}{\rm Mpc/h}\right]-1.27,\hspace{0.3cm}\left(\text{for \,$M_c\geq10^{12}$ M$_{\odot}$/h}\right).
\end{equation}
The amount of total suppression is governed by the factor $B$ (and therefore $M_c$) while the scale of suppression is controlled by $k_g$ which mainly depends on the ejection radius (via the parameter $\eta_b$). The stellar component $S$ is well captured by the simple power law, i.e.
\begin{equation}
S(k|k_s)=1+(k/k_s)^{2}\hspace{1cm}k_s=55\,\,\left[\rm h/Mpc\right].
\end{equation}
The influence of the stellar component $S$ on the power spectrum is negligible below $k\sim 5$ h/Mpc and stays subdominant until $k\sim20$ h/Mpc. As these small scales are not the prime target of this paper, we leave investigations of the dependence between $S$ and the functional form of $f_{\rm cgal}(M)$ to future work.

In Fig.~\ref{fig:PSfit} we show how the fitting function (grey lines) performs with respect to a selection of power spectra with varying $M_{\rm c}$, $\eta_{\rm b}$, and $z$. While the fit does not account for subtleties like the slight change in the total strength of the suppression for varying $\eta_{\rm b}$, it nevertheless provides a good match to all the different power spectra with errors staying well below five percent. This is a remarkable agreement, showing that the baryonic correction of the power spectrum is indeed mainly governed by the two parameters $M_{\rm c}$ and $\eta_{\rm b}$ as well as the redshift.

\subsection{Comparison to previous work}
First analytical estimates of how baryons affect the matter power spectrum have been carried out by Refs.~\citep{White2004,Zhan2004} more than a decade ago. These authors investigated the influence of baryonic cooling \citep{White2004} and hot inter cluster gas \citep{Zhan2004} finding percent effects on the power spectrum. Similar conclusions were obtained a few years later by Refs.~\citep{Jing2006,Rudd2008,Guillet2010} based on hydrodynamical simulations which included radiative cooling and star formation but no AGN feedback. Ref.~\citep{Jing2006} found a power suppression of about two percent beyond $k\sim0.5$ h/Mpc. We approximately recover their result using the fitting function of Eq.~\ref{fit} with the parameters $M_c\sim 2\times10^{12}$ M$_{\odot}$/h and $\eta_b\sim 1.0$. Such a low value for $M_c$ corresponds to a case with basically no ejected gas component, which is in agreement with the lack of AGN feedback in the simulations of Ref.~\citep{Jing2006}.

The first predictions of the matter power spectrum including AGN feedback were performed by Ref.~\citep{vanDaalen2011} about five years ago (see also \citep{Semboloni2011,Harnois-Deraps2015}). Based on simulations capable of reproducing both X-ray and optical observations \citep{Schaye2010,McCarthy2010}, they found a suppression of up to 25 percent with a deviation starting at $k\sim0.3$ h/Mpc. This is a dramatic increase of the effect with respect to earlier investigations and has important implications for the interpretation of current and future weak lensing surveys.

With the {\it baryonic correction model} and using standard model parameters matched to observations, we find a total power suppression roughly similar to the results from Ref.~\citep{vanDaalen2011}. Concerning the fitting function of Eq.~\eqref{fit}, the best match is obtained with parameters $M_c\sim 5\times10^{14}$ M$_{\odot}$/h and $\eta_b\sim 0.4$ pointing towards a strong AGN activity able to push matter out of large clusters. We do not recover the exact shape of the suppression reported by \citep{vanDaalen2011}. In principle, closer agreement could be found by implementing a somewhat shallower ejected gas profile into the {\it baryonic correction model}. However, we want to stress that the shape of the ejected gas profile (and therefore the shape of the power spectrum) crucially depends on the interplay between gas and the AGN which is currently poorly understood\footnote{Recent observational estimations of the AGN feedback energetics from Ref.~\citep{Ruan2015} find considerably larger values than the ones used in feedback implementations of cosmological simulations. This underlines the large uncertainty present in the current modelling of the AGN-gas interplay.}.

Further attempts to parametrise the baryonic effects have been carried out very recently by the Refs.~\citep{Lewandowski2015,Semboloni2013,Mohammed2014,Mead2015,Fedeli2014b} applying analytical methods. While the works of~\citep{Lewandowski2015,Semboloni2013,Fedeli2014b,Mead2015} are all based on results from \citep{vanDaalen2011}, Ref.~\citep{Mohammed2014} exhibits an independent investigation with a parametrisation of baryonic components similar to the one we use. The main differences are that Ref.~\citep{Mohammed2014} built upon the halo model and did neither assume an ejected gas profile (ejected gas is simply added to the linear background perturbations) nor did it include adiabatic DM expansion. They found a total power suppression which is similar to the one we obtain if we neglect adiabatic DM expansion. On the other hand, the largest scale (smallest mode) of suppression found by \citep{Mohammed2014} does not agree with our predictions, most probably because of their simplified modelling of the ejected gas component.

\section{Conclusions}\label{sec:conc}
The influence of baryons on structure formation needs to be quantified in order to extract cosmological information from upcoming full-sky surveys. In this paper we present a new approach to investigate how baryons affect the matter power spectrum based on a modification of $N$-body simulations. The {\it baryonic correction model} assumes every halo to consist of an adiabatically relaxed dark matter profile, a stellar component, as well as a bound and an ejected gas component, which sum up to a modified total halo profile. This profile can be recovered from an initial NFW profile by applying displacement functions to all particles around haloes in a $N$-body simulation.

The main advantage of the {\it baryonic correction model} is that observational measurements of the stellar and gas distributions in haloes can be used to estimate the influence of baryons on the total matter distribution without requiring further input about feedback mechanisms. The effects of single components, such as the ejected gas or central galactic stars, can be studied individually, clarifying the importance of different baryonic components on the total matter density field.

Furthermore, the model simply perturbs the output of $N$-body simulations which are accurate and well controlled calculations of the gravity-only clustering in the nonlinear regime \citep[see e.g][]{Heitmann2010,Reed2013,Schneider2015}. The {\it baryonic correction model} is much more accurate than, for example, the halo model, where both the dark matter power spectrum and the baryonic corrections are estimated analytically. Perturbing $N$-body simulations is a promising approach to produce mock skies for future weak lensing surveys. This is not only true for baryonic corrections, the method is potentially applicable to various deviations from the standard cosmological model, such as the nonlinear effects of massive neutrinos or modified gravity.

In this paper we focus on the power spectrum, which is the prime statistical measure for the large scale structure and contains crucial information about cosmology. Future galaxy and weak-lensing surveys will require precise predictions of the power spectrum in order to be used as cosmological probes. It is therefore crucial to quantify what effects on what scales are expected from the presence of baryons. The {\it baryonic correction model} gives important insights to these problems which we summarise in the following:
\begin{itemize}

\item The most important effect on the power spectrum comes from the gas ejected out of haloes. The {\it ejected gas fraction} per halo mass, $f_{\rm egas}(M)$, regulates how strongly the power spectrum is maximally suppressed with respect to the DM-only case. Based on the {\it baryonic correction model} plus X-ray observations of $f_{\rm egas}(M)$, we predict the power suppression to be 10-25 percent at  scales beyond $k\sim 1$ h/Mpc.

\item Equally important than the ejected gas fraction is the {\it ejected gas radius}, i.e. the typical distance the gas is pushed out of the halo centres. It controls the maximum physical scale (minimum wave mode) where baryons start to affect the power spectrum. Recent measurements of the Sunyaev-Zeldovich signal in large clusters suggest that ejected gas might be around or just outside of the viral radius. Based on these observations the {\it baryonic correction model} predicts the power spectrum to be unaffected by baryons up to wave modes of $k\sim 0.5$ h/Mpc. However, observational uncertainties are large and more data from galaxy groups is required to confirm this constraint.

\item The remaining baryonic components have smaller effects on the power spectrum at medium scales. While the bound X-ray emitting gas only contributes weakly by no more than $\sim3$ percent, the stellar components leads to a strong power increase but only at very small scales beyond $k\sim 5$ h/Mpc.

\end{itemize}
Due to the simple connection between galaxy properties and the power spectrum, it is possible to identify a physically motivated {\it fitting function} for the baryon-induced power suppression. We suggest a two-parameter fit (with additional redshift dependence), one parameter regulating the total amount of suppression and one the maximum scale (minimal wave number) where the suppression exceeds the percent level. The first parameter is linked to the typical mass above (below) which most gas is bound (ejected), while the second is related to the typical ejection radius of the gas. The two parameters have clear physical meaning and capture scale and shape of the power suppression to good accuracy.

Finally, another interesting result of this paper is the rapid convergence rate for ratios of baryonic corrected versus DM-only power spectra. While absolute percent convergence requires simulations with a minimum box-size of $L\sim1000$ Mpc/h and a minimum number of $N\sim4096$ particles per dimension \citep{Heitmann2010,Schneider2015}, relative ratios are converged for $L\sim128$ Mpc/h and $N\sim256$. We argue that resolution and box-size effects appear in both the corrected and the DM-only run and are therefore divided out for relative measures. The rapid convergence rate of ratios is very encouraging, because it means that one high-resolution baseline simulation plus different bias factors based on lower resolution might be sufficient to predict signatures of baryonic physics or alternative cosmologies for next generation surveys.

\section*{Acknowledgements}
We thank Irhsad Mohammed and Joop Schaye for useful discussions. AS is supported by the Synergia project ``Euclid'' from the Swiss National Science Foundation. Simulations were run on the Piz Daint cluster at the Swiss National Supercomputer Centre (CSCS) under the project allocations {\it s511} and {\it s592}.

\begin{appendix}

\section{Appendix}

\subsection{Some more about the truncated NFW profile}\label{app:NFW}
In Sec.~\ref{rhodmo} we have used a generalisation of the NFW profile which is truncated at large radii to assure mass convergence. The profile is given in Eq.~\eqref{rhonfw} but for the sake of clarity, here it is again
\begin{equation}
\rho_{\rm nfw}(x.\tau)\equiv \frac{\rho_0}{x(1+x)^2}\frac{1}{(1+(x/\tau)^2)^2}\,,
\end{equation}
with $x=r/r_s$, $\tau=r_{\rm tr}/r_{s}$ and $r_s<r_{\rm 200}<r_{\rm tr}$ for scale, virial, and truncation radius. The profile has been introduced and discussed in the Refs.~\citep{Baltz2009,Oguri2011}. The corresponding mass profile can be integrated analytically and is given by
\begin{equation}
\begin{split}\label{Mnfw}
M_{\rm nfw}(x,\tau)=&4\pi\rho_0 \left(\frac{r_{\rm 200}}{c}\right)^3\,m_{\rm nfw}(x,\tau)\,,\\
m_{\rm nfw}(x,\tau)\equiv&\frac{\tau^2}{2(1+\tau^2)^3}\left\{ \frac{x\left[x-2\tau^6+x(1-3x)\tau^4 +x^2 +2(1+x-x^2)\tau^2\right]}{(1+x)(\tau^2+x^2)}\right.\\
& \left.+\tau(6\tau^2-2)\arctan\left(\frac{x}{\tau}\right)+\tau^2(\tau^2-3)\ln\left[\frac{\tau^2(1+x)^2}{\tau^2+x^2}\right]\right\}
\end{split}
\end{equation} 
With the mass profile at hand, we can easily define viral mass, and total mass
\begin{equation}\label{Mvir}
\begin{split}
M_{\rm 200}&\equiv M_{\rm nfw}(c,\tau)=4\pi\rho_0 \left(\frac{r_{\rm 200}}{c}\right)^3\,m_{\rm nfw}(c,\tau),\\
M_{\rm tot}&\equiv M_{\rm nfw}(\infty,\tau)=4\pi\rho_0 \left(\frac{r_{\rm vir}}{c}\right)^3m_{\rm nfw}(\infty,\tau).
\end{split}
\end{equation}
It is straight-forward to show that in the limit $x\rightarrow \infty$ the second part of Eq.~\eqref{Mnfw} becomes
\begin{equation}
m_{\rm nfw}(\infty,\tau)=\frac{2\tau^4(\tau^2-3)\ln(\tau)-\tau^2(3\tau^2-1)(\tau^2-\pi\tau+1)}{2(1+\tau^2)^3}.
\end{equation}
Since the viral mass is defined as $M_{\rm 200}=4\pi\Delta_{\rm 200}\rho_cr_{\rm 200}^3/3$, we can rewrite the truncated NFW density and mass profiles with respect to the background density $\bar\rho$ instead of $\rho_0$, i.e.,
\begin{subequations}
\begin{align}
\rho_{\rm nfw}(x.\tau)&=\frac{4\pi\Delta_{\rm 200}c^3}{3\,m_{\rm nfw}(c,\tau)}\frac{1}{x(1+x)^2(1+(x/\tau)^2)^2}\,,\\
M_{\rm nfw}(x,\tau)&=\frac{4\pi}{3}\Delta_{\rm 200}\bar\rho r_{\rm 200}^3\frac{m_{\rm nfw}(x,\tau)}{m_{\rm nfw}(c,\tau)}\,.
\end{align}
\end{subequations}

\begin{figure}[tbp]
\center{
\includegraphics[width=0.47\textwidth,trim={1cm 1cm 3cm 0cm}]{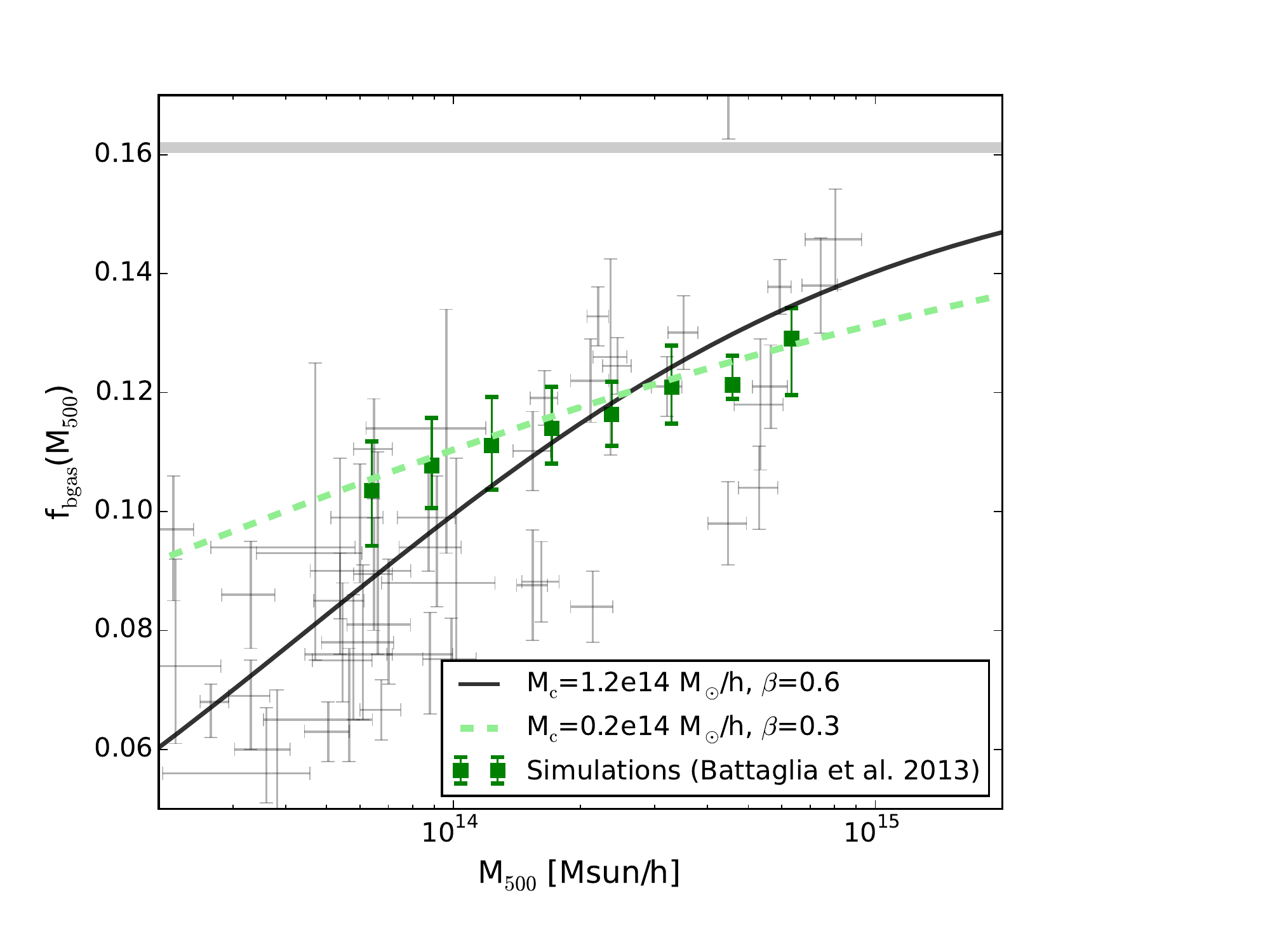}
\includegraphics[width=0.47\textwidth,trim={1cm 1cm 3cm 0cm}]{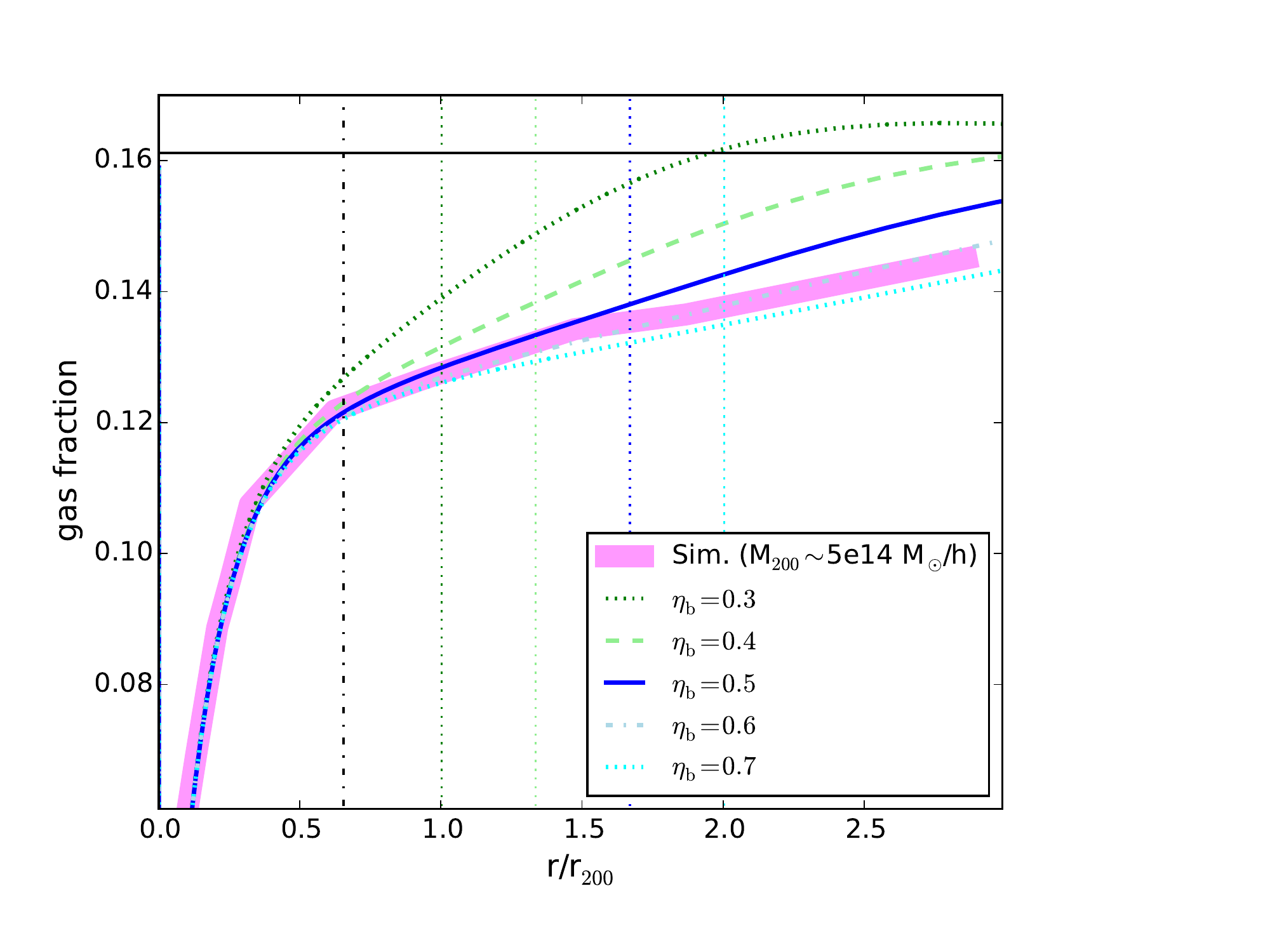}\\
\includegraphics[width=0.47\textwidth,trim={1cm 1cm 3cm 0cm}]{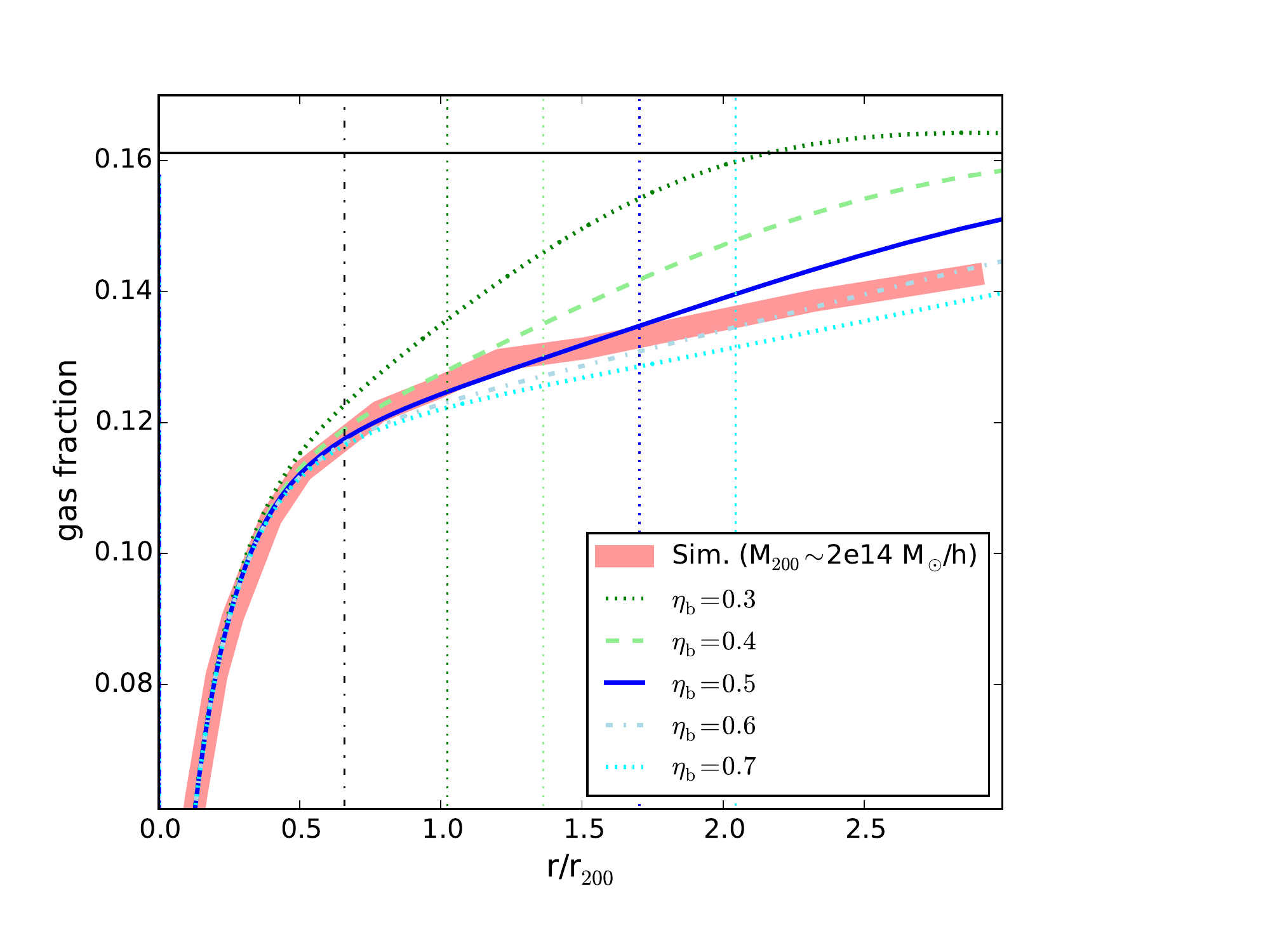}
\includegraphics[width=0.47\textwidth,trim={1cm 1cm 3cm 0cm}]{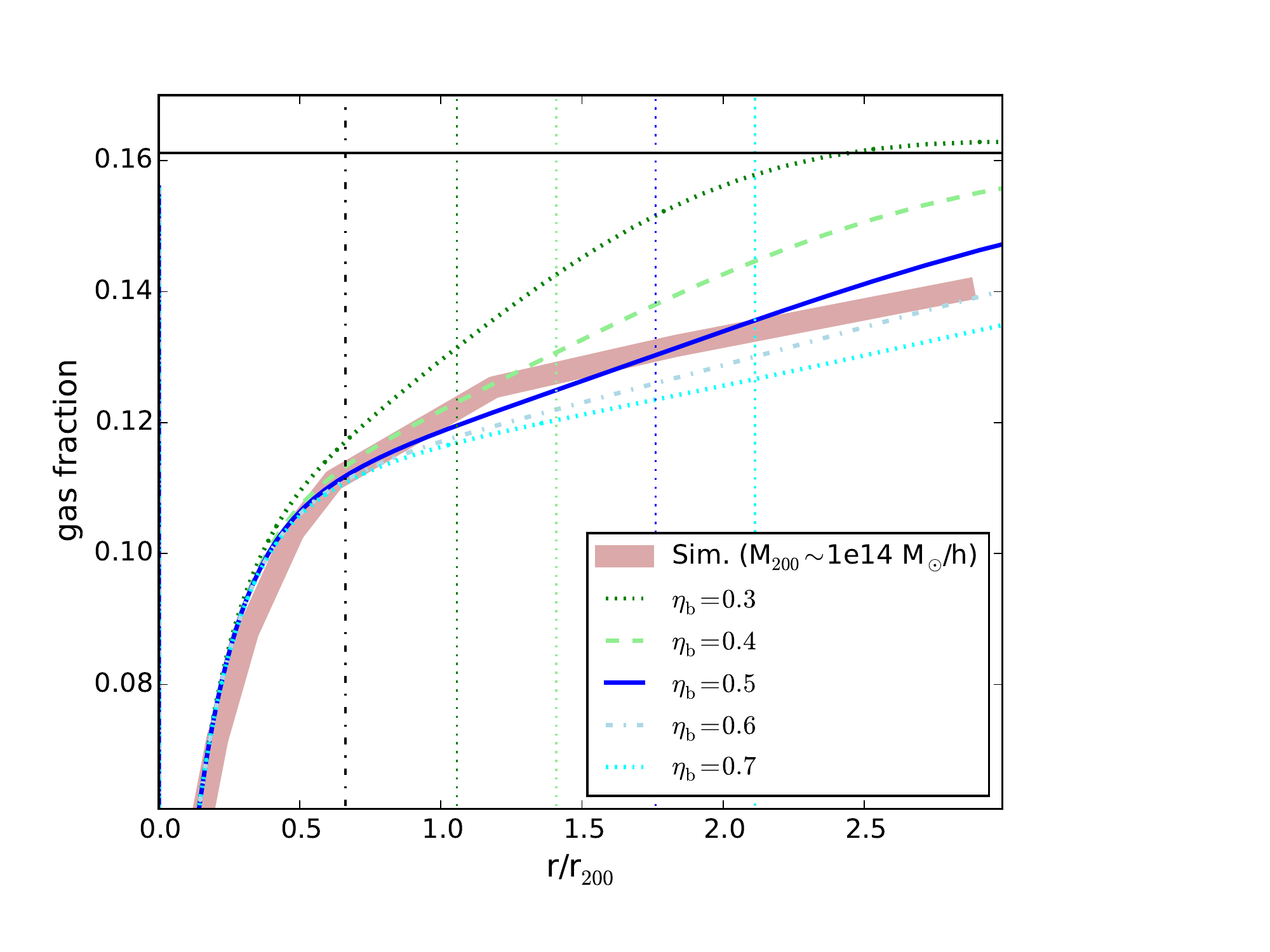}
\caption{\label{fig:sim}Upper left: hot gas fraction at $M_{500}$ from the simulations of Ref.~\citep{Battaglia2013} (green symbols) and from observations (grey symbols, same as in Fig.~\ref{fig:PS_Mcbeta}). Other panels: Comparison between the gas profile from the {\it baryonic correction model} and stacked gas profiles from the simulations (thick coloured lines) for different cluster masses.}}
\end{figure}

\subsection{Comparison with hydrodynamical simulations}\label{app:sim}
In the main text of this paper we deliberately omitted to show results from hydrodynamical simulations because the {\it baryonic correction model} is supposed to provide an alternative measure of the power spectrum solely based on observations of gas distribution. A comparison with simulations is nevertheless helpful in order to verify the viability of the model parameters. Throughout the paper we have already mentioned several references where parts of the {\it baryonic correction model} have been tested with simulations. For example, the authors of Ref.~\citep{Teyssier2011} have found good agreement between the baryonic back-reaction model (given by Eq.~\ref{ACmodel}) and their simulations. In Refs.~\citep{Martizzi2013,Mohammed2014} it was shown that the stellar and gaseous profiles inside haloes (given by Eqs.~\ref{rhohga} and \ref{rhocga}) are able to reproduce results from cluster zoom simulations. The model for the gas outside of the viral radius (i.e the ejected gas profile of Eq.~\ref{rhoega}), however, has not been compared to simulations so far.

In this appendix we use results from hydrodynamical simulations of Ref.~\citep{Battaglia2013} to compare gas fractions in and outside the viral radius. This is an important consistency test for the {\it baryonic correction model} as the gas fraction directly depends on the underlying bound and ejected gas profiles. The simulations of Ref.~\citep{Battaglia2013} were run to study galaxy groups and clusters as well as the intra-cluster medium. They used an AGN recipe where the black hole accretion is coupled to the star formation rate. This simple model has been shown to produce identical results in terms of gas distribution than more elaborated AGN models \citep{Battaglia2010} and is ideal for simulating large boxes with a large clusters sample.

In the top-left panel of Fig.~\ref{fig:sim} we show the simulated gas fraction at $M_{500}$ (green symbols in top left panel) which is in good agreement with observations (grey symbols, same as in Fig.~\ref{fig:fhga}) albeit with a somewhat shallower mass dependence (green dashed line) than the best fit to  the observations (grey solid line, see Sec.~\ref{sec:frac}). In the remaining panels we illustrate the mean gas profiles of clusters from the simulations (thick coloured lines) together with the predictions from the {\it baryonic correction model} depending on the value of $\eta_b$ (see Eq.~\ref{rcr2}). Different panels refer to different cluster masses from $10^{14}$ to $5\times 10^{15}$ M$_\odot$/h.

Fig.~\ref{fig:sim} shows that the simulated gas profiles from Ref.~\citep{Battaglia2013} agree well with the model predictions, which validates our assumptions regarding the density profiles of the ejected gas component. The best agreement is found with the parameter choice of $\eta_b\sim 0.6$. This is in some tension with the observations illustrated in Fig.~\ref{fig:PS_rej} of the main text, which seem to favour smaller values of $\eta_b\lesssim 0.4$.

\end{appendix}

\end{document}